\documentclass[fleqn,usenatbib]{mnras}
\usepackage{newtxtext,newtxmath}
\usepackage[T1]{fontenc}

\DeclareRobustCommand{\VAN}[3]{#2}
\let\VANthebibliography\thebibliography
\def\thebibliography{\DeclareRobustCommand{\VAN}[3]{##3}\VANthebibliography}

\usepackage{graphicx}
\usepackage{afterpage}
\usepackage{amsmath}

\newcommand{\msun}{{\rm M}_\odot}
\newcommand{\msunyr}{{\rm M}_\odot\,{\rm yr}^{-1}}
\newcommand{\cc}{{\rm cm^{-3}}}
\newcommand{\kms}{{\rm km\,s^{-1}}}

\newcommand{\nth}{n_{\rm th}}
\newcommand{\tth}{t_{\rm th}}

\newcommand{\vsv}{v_{\rm SV}}
\newcommand{\sigmasv}{\sigma_{\rm SV}}
\newcommand{\vsvsigma}{\vsv/\sigmasv}
\newcommand{\Rvir}{R_{\rm v}}
\newcommand{\Mvir}{M_{\rm v}}
\newcommand{\Tvir}{T_{\rm v}}

\newcommand{\Ncore}{N_{\rm c}}

\newcommand{\Nstar}{N_{\rm s}}
\newcommand{\Nstarcloud}{N_{\rm s/c,max}}
\newcommand{\Mstar}{M_{\rm s}}
\newcommand{\Mstartot}{M_{\rm s,tot}}
\newcommand{\Mstarmax}{M_{\rm s,max}}

\newcommand{\dmdtin}{\dot{M}_{\rm in}}
\newcommand{\Mcrit}{M_{\rm crit}}

\newcommand{\nmodel}{n_{\rm model}}

\graphicspath{{./}{figures/}}

\defcitealias{Hirano2023FSC1}{Paper~I}
\defcitealias{Hirano2025FSC2}{Paper~II}

\title[First star clusters - III]{Formation of first star clusters under the supersonic gas flow -- III. \\ Environmental trends and halo-to-halo scatter in the Pop III IMF}

\author[S. Hirano et al.]{
Shingo Hirano,$^{1,2}$\thanks{E-mail: shingo-hirano@kanagawa-u.ac.jp}
Yusuke Sakai$^{2}$ and 
Hideyuki Umeda$^{2}$\\
$^{1}$Department of Applied Physics, Faculty of Engineering, Kanagawa University, Kanagawa 221-0802, Japan\\
$^{2}$Department of Astronomy, School of Science, University of Tokyo, Tokyo 113-0033, Japan
}

\date{Accepted 2026 March 31. Received 2026 March 20; in original form 2026 February 5}
\pubyear{2026}

\begin{document}
\label{firstpage}
\pagerange{\pageref{firstpage}--\pageref{lastpage}}
\maketitle

\begin{abstract}
The first generations of stars ionised and enriched their host galaxies and seeded the growth of massive black holes.
Models often assume that Pop~III stellar masses in different minihaloes are stochastic realisations of a single universal initial mass function (IMF).
We use 138 cosmological zoom-in hydrodynamics simulations to test this assumption and to map the first-star IMF across redshift, halo mass, and baryon-dark matter streaming velocity (SV).
We construct a dense-cloud merger tree and assign first-star masses by mapping the radial gas accretion-rate profile to stellar mass, yielding per-halo stellar mass functions without imposing any \emph{a priori} IMF.
The high-mass tail and multiplicity increase systematically with redshift, halo mass, and SV.
Low-mass, low-SV haloes form only one or a few first stars, whereas massive, high-SV haloes host rich first star clusters and commonly produce very massive ($\gtrsim10^3$--$10^4\,{\rm M}_\odot$) first stars.
Even in a fixed environment, halo-to-halo scatter remains substantial.
Nevertheless, group-averaged IMFs converge to well-defined forms, ruling out a single universal IMF at the halo level across the range of environments probed here.
Mapping our seeds onto the redshift--mass plane, we show that high-SV and massive haloes preferentially populate the heavy-seed regime relevant to luminous high-redshift sources.
At the same time, low-SV environments are consistent with single/few-event enrichment signatures in metal-poor stars.
Our results deliver a practical, physically motivated prescription for per-halo IMF.
\end{abstract}

\begin{keywords}
methods: numerical --
dark ages, reionization, first stars --
stars: Population III --
stars: formation --
stars: black holes
\end{keywords}


\section{Introduction}
\label{sec:intro}

The onset of cosmic dawn, the formation of the first generation of stars (Population~III stars; Pop~III stars) and galaxies in the early Universe, is now being probed directly through observations.
First, recent deep surveys with the \textit{James Webb Space Telescope} (\textit{JWST}) have revealed galaxies with secure spectroscopic redshifts $z>14$ \citep[e.g.][]{Carniani2024_z14,Naidu2025_Mom-z14}, as well as chemically evolved systems whose metal and carbon abundances indicate prior enrichment by first (metal-free) stars \citep[e.g.][]{Scholtz2025_GS-z11-1}.
The inferred number densities of such very high-redshift galaxies appear to exceed the expectations from pre-\textit{JWST} ``standard'' galaxy-formation models calibrated primarily at lower redshifts, and some systems already show signatures suggestive of very massive Pop~III sources \citep[e.g.][]{Maiolino2024}.
These results motivate models in which first stars and their enrichment episodes begin at higher redshift than previously assumed.

Second, complementary observations indicate that pristine or nearly pristine star formation may persist to lower redshifts.
Strongly lensed candidates such as LAP1 at $z\simeq6.64$ \citep[e.g.][]{Vanzella2023} and LAP1-B \citep{Nakajima2025_LAP1-B,Visbal2025_LAP1-B} motivate scenarios in which metal-free star formation is delayed, re-triggered, or confined to late-time pockets.
Consistently, large-scale simulations find that Pop~III star formation can linger down to the epoch of reionization (EoR) under favourable conditions \citep[e.g.][]{Xu2016,Zier2025}, providing a theoretical context for lensed pristine candidates.

Third, the near-field fossil record in extremely metal-poor (EMP) stars provides an additional line of evidence \citep[e.g.][]{FrebelMorris2015}.
EMP abundance patterns constrain the multiplicity and mass scale of early enrichment events \citep[e.g.][]{Aoki2014,Bessell2015}, and enrichment multiplicity is environmentally dependent \citep[e.g.][]{Hartwig2018,Hartwig2023}.
In this near-field regime, the scatter and high-mass tail of the Pop~III initial mass functions (IMF) become as important as the mean IMF, and improved diagnostics continue to sharpen constraints on very massive Pop~III explosions \citep[e.g.][]{Vanni2024}.
Together, the high-$z$, low-$z$, and near-field lines of evidence indicate that Pop~III star formation and enrichment must be modelled over an extended redshift interval and across diverse environments.

To interpret this diverse set of observations, we require physically motivated models of the first-star IMF that remain applicable across a broad range of environments and redshifts.
High-redshift spectral modelling and line-ratio diagnostics indicate that very massive Pop~III stars may be needed in some systems, such as nitrogen-rich galaxies where $10^{3}$--$10^{4}\,\msun$ first stars have been suggested \citep{Nandal2025_1e3-1e4MsunPopIII}.
Recent \textit{JWST} surveys have also revealed a growing population of compact ``Little Red Dots'' \citep[LRDs; e.g.][]{Hviding2025}, sharpening the need for physically grounded pathways to early massive black hole (BH) seeds and motivating forward-modelling efforts that connect embedded active galactic nuclei (AGN) observables to their underlying seed populations \citep[e.g.][]{Volonteri2025}.
A coherent IMF description must therefore span at least $z\sim30$ down to $z\sim3$.
It must also encompass a range of environments, including LW-irradiated regions, ionized and X-ray heated gas, and dynamically perturbed haloes.

The formation of first stars has long been studied with numerical simulations \citep[for recent reviews, see][]{KlessenGlover2023}.
To interface such simulations with semi-analytic and galaxy-formation models, sub-grid prescriptions are being developed that encode Pop~III formation in terms of a few effective parameters, such as the critical halo mass $\Mcrit$, the formation efficiency, and an assumed IMF \citep[e.g.][]{Kulkarni2021,Hazlett2025,Gurian2024}.
Environmental effects on $\Mcrit$ and early cloud properties, including LW radiation and baryon-dark matter (DM) streaming velocity (SV), have been quantified in dedicated simulations and semi-analytic calibrations \citep[e.g.][]{Greif2011sv,Stacy2011,SkinnerWise2020,Lenoble2024}.
In contrast, the redshift dependence of the Pop~III IMF and the efficiency of forming multiple stars or clusters within a single halo remain comparatively poorly constrained in cosmological settings.
For example, \citet{Hirano2015} demonstrated that the characteristic first-star mass can evolve significantly with redshift, but such studies have not yet been systematically extended to lower redshifts and to a wider range of halo environments.
Likewise, multiple Jeans-unstable clouds may form within a single halo, for example, under strong LW irradiation or large streaming velocities \citep[e.g.][]{Hirano2018}.
Such conditions can give rise to first star clusters (FSCs), but have not been explored in a systematic, cosmological context.

In this context, many semi-analytic and galaxy-formation models adopt a convenient null hypothesis: Pop~III stellar masses in different minihaloes are stochastic realizations of a single, universal IMF, loosely analogous to the local ``standard'' IMF \citep{Salpeter1955,Kroupa2001,Chabrier2005}.
We, on the other hand, treat IMF universality as a working hypothesis rather than a prior and test it directly.
Specifically, we map Jeans-unstable cloud properties to stellar masses (rather than sampling an assumed IMF), construct per-halo Pop~III mass functions, and quantify their scatter as a function of $(z, \Mvir, \vsv)$.
We then assess how non-universality impacts the interpretation of both high-redshift observations and near-field chemical constraints.

This study is the third paper in the FSC series.
\citetalias{Hirano2023FSC1} \citep{Hirano2023FSC1} and \citetalias{Hirano2025FSC2} \citep{Hirano2025FSC2} established the simulation suite and the conditions for FSC formation under streaming.
We introduce a cloud-scale merger-tree framework and a per-halo IMF prescription suitable for sub-grid models.
Recent work continues to refine Pop~III formation thresholds and the role of streaming \citep[e.g.][]{Nebrin2023,Chen2025}, providing context for interpreting our group-dependent IMF trends.

Our main goal is to bridge the gap between halo-scale and cloud-scale descriptions of first-star formation by resolving Jeans-unstable gas clouds down to densities of $n\sim10^{6}\,\cc$ and by following their subsequent evolution and mergers within their host haloes.
We construct a merger tree at the scale of dense gas clouds instead of at the halo scale alone.
We then quantify how halo assembly history and baryon-DM streaming affect the cloud multiplicity, cloud mergers, and the resulting first-star population.
We focus on the larger-scale processes (halo assembly, the formation and growth of Jeans-unstable clouds, and cloud mergers) that set the stage for FSC formation.
Detailed protostellar physics and small-scale disc fragmentation will be refined by future ultra-high-resolution simulations, but are beyond the scope of the present work.

This paper is organized as follows.
In Section~\ref{sec:methods_sim}, we describe the cosmological simulations and the additional models introduced to probe higher-redshift, more massive haloes.
Section~\ref{sec:methods_tree} presents our methodology for identifying Jeans-unstable gas clouds, constructing their merger trees, and modelling first-star formation.
In Section~\ref{sec:res}, we show the resulting statistics of cloud formation, mergers, and first-star populations across different streaming velocities and halo environments.
Section~\ref{sec:dis} discusses the implications of our findings for the Pop~III IMF and for the assembly of early BHs and galaxies.
Finally, Section~\ref{sec:con} summarizes our main conclusions and outlines directions for future work.

\section{Cosmological Simulations}
\label{sec:methods_sim}

\begin{table}
\centering
\caption{Parameters of the cosmological simulations}
\label{table:params}
\begin{tabular}{lll}
\hline
Parameter & Value & Model identifier \\
\hline
\multicolumn{3}{l}{({\it $\Lambda$CDM cosmology})} \\
$\Omega_{\rm m}$ & 0.31 & \\
$\Omega_{\rm b}$ & 0.048 & \\
$\Omega_\Lambda$ & 0.69 & \\
$H_0$ & $68\,\kms\,{\rm Mpc}^{-1}$ & \\
$\sigma_8$ & \{0.83, 1.0, 1.1, 1.2\} & \{S08, S10, S11, S12\} \\
$n_{\rm s}$ & 0.96 & \\
& & \\
\multicolumn{2}{l}{({\it Cosmological initial condition})} & \{I01 -- I20\} \\
$z_{\rm ini}$ & 499 & \\
$L_{\rm base}$ & $10\,h^{-1}$\,cMpc & \\
$m_{\rm dm,base}$ & $7.966\times10^5\,\msun$ & \\
$m_{\rm gas,base}$ & $1.460\times10^5\,\msun$ & \\
$L_{\rm zoom}$ & $0.3\,h^{-1}$\,cMpc & \\
$m_{\rm dm,zoom}$ & $24.31\,\msun$ & \\
$m_{\rm gas,zoom}$ & $4.454\,\msun$ & \\
& & \\
\multicolumn{3}{l}{({\it Initial streaming velocity})} \\
$\vsvsigma$ & \{0, 1, 1.5, 2, 2.5, 3\} & \{V00, V10, V15, V20, V25, V30\} \\
$\sigmasv(z_{\rm ini})$ & $13.76\,\kms$ & \\
& & \\
\multicolumn{3}{l}{({\it Cosmological simulation})} \\
Chemistry & \multicolumn{2}{l}{e$^-$, H, H$^+$, H$^-$, H$_2$, H$_2^+$} \\
& \multicolumn{2}{l}{D, D$^+$, HD, HD$^+$, HD$^-$, He, He$^+$, He$^{++}$} \\
$m_{\rm dm,min}$ & $0.1439\,\msun$ & \\
$m_{\rm gas,min}$ & $0.02636\,\msun$ & \\
$L_{\rm Jeans}$ & $>15 L_{\rm hsml}$ & \\
$\nth$ & $10^6\,\cc$ & \\
\hline
\end{tabular}
\begin{flushleft}
{\it Notes.} We adopt the $\Lambda$CDM cosmology \citep{PLANCK2018}.
\end{flushleft}
\end{table}

Our analysis is based on cosmological smoothed-particle hydrodynamics (SPH) simulations of the first stars.
The simulation suite and numerical methodology closely follow our previous work on the formation of FSCs under the supersonic gas flow \citetalias{Hirano2025FSC2}.
For completeness, the key parameters of the simulations are summarised in Table~\ref{table:params}.
Here, we provide only a brief overview of the setup and emphasise the aspects most relevant to the present study.

We adopt a flat $\Lambda$ cold dark matter ($\Lambda$CDM) cosmology consistent with the \citet{PLANCK2018} results, with matter and baryon density parameters, Hubble constant, and primordial power spectrum parameters as listed in Table~\ref{table:params}. 
Cosmological initial conditions are generated at $z_{\rm ini}=499$ in a periodic box of side length $L_{\rm base}=10\,h^{-1}$\,cMpc, from which we identify 20 representative first-star-forming haloes.
Around each target halo, we construct a nested zoom-in region of size $L_{\rm zoom}=0.3\,h^{-1}\,$\,cMpc, realised with DM and gas particle masses $m_{\rm dm,zoom}$ and $m_{\rm gas,zoom}$.
The zoom-in simulations include a primordial non-equilibrium chemical network (H, He, D, and their ions and molecules) and associated cooling/heating processes, allowing us to follow the collapse of metal-free gas clouds from the cosmological background down to high densities.
We explore baryon-DM streaming velocities spanning $\vsvsigma=0$--3.

To capture the long-term evolution of star-forming clouds in a numerically stable manner, we adopt the same ``opaque-core'' treatment as in our earlier work.
Once the central gas reaches the threshold density $\nth=10^6\,\cc$, we suppress further radiative cooling above this density.
We then continue to follow the dynamics and accretion for an additional 2\,Myr (physical time).
Because 2\,Myr is shorter than the typical lifetime of a first star \citep[e.g.][]{Schaerer2002}, supernova feedback does not affect our haloes during this period.
We therefore neglect SN-driven dynamical and chemical feedback in the present analysis.
The present study uses the resulting SPH particle data over this 2\,Myr window with a time spacing of $10^4$\,yr (Section~\ref{sec:methods_tree}).

\subsection{Additional models}
\label{sec:methods_additional-models}

The bulk of our simulation data is consistent with our previous work, in which we analysed 120 cosmological zoom-in simulations (Table~\ref{table:120models}) spanning a range of initial streaming velocities between baryons and DM.
The streaming velocity, normalised by its root-mean-square value at recombination, takes the values $\vsvsigma=\{0, 1, 1.5, 2, 2.5, 3\}$, corresponding to the model identifiers V00, V10, V15, V20, V25, and V30, and the 20 cosmological initial conditions are denoted by I01--I20.
The baryon-DM relative velocity originates at recombination and remains coherent on comoving Mpc scales.
This large-scale field modulates small-scale structure formation by delaying collapse, reducing baryon fractions in minihaloes, and altering the thermodynamic and turbulent state of primordial gas \citep{Tseliakhovich2010, Fialkov2014}.

In the present work, we extend the original suite by varying the amplitude of the linear matter power spectrum via $\sigma_8$, the rms amplitude of linear matter fluctuations on $8\,h^{-1}$\,Mpc (Table~\ref{table:new18models}).
Our fiducial simulations adopt $\sigma_8=0.83$, consistent with \cite{PLANCK2018}.
We additionally generate initial conditions with $\sigma_8 = \{0.83, 1.0, 1.1, 1.2\}$, labelled S08, S10, S11, and S12.
Increasing $\sigma_8$ shifts the collapse of rare haloes to earlier times and yields more massive first-star-forming haloes at higher redshift.
This allows us to probe the regime relevant to very high-redshift compact galaxies and over-massive AGN without altering the simulation microphysics.

We adopt the classification introduced in \citetalias{Hirano2025FSC2}, which groups haloes into Classes {\it High}, {\it Middle}, and {\it Low} in the joint $(z, \Mvir)$ plane.
Here, $z$ denotes the redshift at which the gas first reaches the density threshold for cloud identification, and $\Mvir$ is the corresponding virial mass of the host halo.
Following this scheme, we select three representative cosmological initial conditions (I02, I04, and I16) as prototypes of these classes.
For each of these three initial conditions, we recompute a set of simulations.
We vary $\sigma_8$ over three enhanced values (S10, S11, and S12) and adopt two extreme streaming velocities (V00 and V30, i.e. zero and $3\sigma$ streaming).
This yields $3 \times 3 \times 2 = 18$ additional models in total.
Because the three initial conditions represent haloes forming at different epochs and masses, and the enhanced $\sigma_8$ values further shift the collapse times, the combined sample of 18 models spans first-star-forming haloes over a wide redshift range, $z \sim 50$ down to $z \sim 15$ (Figure~\ref{fig:z-Mvir_SV120+18}).

These 18 additional models are constructed from the three representative initial conditions (I02, I04, and I16), combined with three enhanced $\sigma_8$ values (S10, S11, and S12) and two streaming-velocity choices (V00 and V30), while keeping all other cosmological and numerical parameters identical to those of the fiducial runs.

For clarity, we adopt a unified model naming convention in which each simulation is identified by a triplet, IiiSssVvv, where Iii specifies the cosmological initial condition
(I01--I20), Sss indicates the value of $\sigma_8$ (S08, S10, S11, and S12), and Vvv encodes the streaming velocity (V00--V30).
The previously published simulations correspond to S08, while the 18 additional models all have S10, S11, and S12. In the subsequent analysis, we use all models as inputs to the merger-tree construction described in Section~\ref{sec:methods_tree}.
However, unless otherwise noted, the statistical results and group-averaged trends (Tables~\ref{table:group_class-vs_mean}--\ref{table:group_class-vs_imf} and Figures~\ref{fig:Vsv-Ncloud-Nstar-rate}--\ref{fig:imf-dist}) are computed from the original 120-model S08 suite only.
The additional 18 models are designed to probe rare, high-redshift, more massive haloes and therefore occupy a distinct region in the $(z, \Mvir)$ plane (Figure~\ref{fig:z-Mvir_SV120+18}), making them unsuitable for inclusion in the group averages defined for the S08 sample.
We use these additional models primarily to extend the $(z, \Mvir)$ coverage and to provide high-$z$ context in Figure~\ref{fig:z-Mhalo-Mbh}.

\section{Merger tree analysis}
\label{sec:methods_tree}

We analyse a cosmological SPH simulation that is continued for a total duration of 2\,Myr from a reference time $\tth=0$\,yr.
During this interval, we output snapshots at uniform intervals of $10^4$\,yr, yielding 201 snapshots in total (labelled 0 to 200).
All post-processing described below is performed on these snapshots and their interrelationships.

Our post-processing consists of three steps: (1) identifying dense gas clouds (hereafter ``nodes'') in each snapshot, (2) constructing a merger tree of these nodes across the 201 snapshots, and (3) applying a star-formation recipe along the merger tree to model the formation of the first stars.

\subsection{Identification of dense gas clouds (nodes)}

For each snapshot, we first select only the high-density SPH gas particles with number density $n > \nth = 10^6\,\cc$.
Using this subset, we identify self-gravitating dense clouds (nodes) via a friends-of-friends (FOF)-like neighbour search in configuration space:
\begin{enumerate}
  \item We pick a particle that has not yet been assigned to any node and initialise a new node with this particle.

  \item We search for all high-density particles within a linking length $R_{\rm link} = 0.25$\,pc from any particle already assigned to the node, and add all such neighbours to the same node.

  \item For every newly added particle, we repeat the neighbour search described above, iterating the procedure until no further particles satisfy the distance criterion with respect to any particle in the node.

  \item We then select another unassigned high-density particle, initialise a new node, and repeat the same procedure. This continues until all particles with $n > \nth$ are assigned to some node.
\end{enumerate}
After this clustering step, we compute the total gas mass $M_{\rm node}$ for each node by adding the masses of its SPH particles.
Nodes with $M_{\rm node} \le 25\,\msun$ (corresponding to $\sim\!1000$ SPH particles in our simulation) are discarded and excluded from the subsequent merger-tree and star-formation analysis.
This mass cut removes small, transient overdensities unlikely to host the first star formation during the period of interest.

\subsection{Construction of the node merger tree}

Using the set of nodes identified in each snapshot, we construct a merger tree over all 201 snapshots.
We link nodes across consecutive snapshots by tracking their constituent SPH particles.
Let node $A$ be a node in snapshot $i$ and node $B$ be a node in snapshot $i+1$.
We count the number of SPH particles shared by $A$ and $B$, and define the overlap fraction (with respect to $A$) as
\begin{equation}
  f_{A\rightarrow B} \equiv \frac{N_{\rm shared}(A,B)}{N_{\rm part}(A)} \ .
\end{equation}
We define the descendant of $A$ as the node $B$ in snapshot $i+1$ with $f_{A\rightarrow B} \ge 0.5$, if such a node exists.
This threshold requires that at least half of $A$'s mass (in SPH particle count) is inherited by the descendant.
Repeating this from snapshot 0 to 201 yields merger trees (evolutionary branches) that trace dense-cloud assembly over the full 2\,Myr interval.
We record a ``cloud merger'' whenever a node in snapshot $i+1$ has two or more distinct progenitors in snapshot $i$ under the same criterion.

\subsection{Star formation recipe}

We next apply a first-star-formation recipe to the node-merger tree to model the formation of the first stars within dense gas clouds.

\subsubsection{Free-fall time and merger-induced delay}

At the threshold density $\nth = 10^6\,\cc$ and $\rho = \mu m_{\rm H} \nth$ (with mean molecular weight $\mu \simeq 1.22$), the free-fall time is
\begin{equation}
  t_{\rm ff} = \sqrt{\frac{3 \pi}{32 G \rho}} \simeq 0.05\,{\rm Myr} \ ,
\end{equation}
which corresponds to $N_{\rm ff} = 5$ snapshots in our time sampling.
We assume that cloud mergers temporarily disrupt or delay gravitational collapse and star formation within the dense gas.

Operationally, for each node at snapshot i, we look back $N_{\rm ff} = 5$ previous snapshots along its merger-tree branch.
If any cloud merger occurred within this interval (i.e. a node with multiple progenitors), we suppress star formation in the current node.
Star formation can therefore proceed only after a dynamically quiet phase of at least one free-fall time at $\nth$.

\subsubsection{Criterion for the star formation}

We now scan the merger tree in chronological order, from earlier to later snapshots.
For each node in snapshot $i$, we evaluate the following conditions:
\begin{enumerate}
  \item No cloud merger has occurred in the last $N_{\rm ff} = 5$ snapshots along its evolutionary branch (as described above).
  \item Neither the node itself nor any of its progenitors in earlier snapshots has previously hosted first star formation.
\end{enumerate}
If both conditions are satisfied, we assume that a first star forms within the node at snapshot $i$.
The node is then flagged as having experienced first star formation.

\subsubsection{Assignment of the stellar mass}
\label{sec:methods:mapping}

We estimate the mass of each first star by converting the gas accretion rate onto the Jeans-unstable cloud into a final stellar mass.
For each cloud (merger-tree node), we evaluate the inflow rate from a spherically averaged radial profile\footnote{At this Jeans radius, typically on sub-parsec scales, the cloud already forms a self-gravitating bound structure, so the influence of larger-scale asymmetry on the spherically averaged inflow profile is expected to be minor.} centered on the density peak at the threshold stage, $n_{\rm th}=10^6\,\cc$.
From this profile, we identify the Jeans radius as the radius within which the enclosed gas mass exceeds the local Jeans mass, and define $\dmdtin$ as the gas inflow rate measured at that radius.
Thus, the adopted $\dmdtin$ represents the characteristic cloud-scale accretion rate associated with the gravitationally unstable region used for mass assignment.

To do so, we follow the analytic framework developed by \citet{Toyouchi2023} and refined by \citet{Liu2024_PopIII}, who constructed a unified model for the final mass of very massive metal-free stars as a function of the inflow rate ($\dmdtin$).
In their formulation, the final stellar mass is determined by the most restrictive among several physical limits.
These include a feedback-limited mass ($M_{\rm s,f}$) at which photoionization-driven outflows halt accretion, and a bloating-mode mass ($M_{\rm s,b}$) relevant at very high inflow rates where the protostar remains swollen and cool.
The model also accounts for a lifetime-limited mass ($M_{\rm s,l}$), a general-relativistic-instability limit ($M_{\rm s,g}$), and, when applicable, a gas-supply limit ($M_{\rm c}$).
The full model combines these components as
\begin{equation}
\Mstar(\dmdtin) = \min \Bigl[ \max\bigl(M_{\rm s,b}, M_{\rm s,f}\bigr), M_{\rm s,l}, M_{\rm s,g}, M_{\rm c} \Bigr] \ ,
\end{equation}
and reproduces the results of detailed protostellar-evolution and radiation-hydrodynamics simulations over a wide range of accretion rates.

Implementing the full multi-branch model for all clouds in our 138 haloes would add considerable complexity, while our merger-tree analysis mainly requires a robust mapping between ($\dmdtin$) and the characteristic stellar mass.
We therefore adopt a simplified version of the \citet{Liu2024_PopIII} model that reproduces their fiducial relation (their Figure~1) to good accuracy over the accretion-rate range relevant for our clouds, $\dmdtin \sim 10^{-4}-10^{-1}\,\msunyr$.
We approximate the feedback-limited branch with the \citet{Hirano2014}, based fit
\begin{equation}
M_{\rm s,f}(\dmdtin) = 100\,\msun \left( \frac{\dmdtin} {2.8\times 10^{-3}\,\msunyr} \right)^{0.8} \ ,
\label{eq:Mstar_dMdt_Hirano14}
\end{equation}
which captures the quasi-linear increase of the final mass with accretion rate found in two-dimensional radiation-hydrodynamics simulations.
For higher inflow rates, we include a bloating-mode branch calibrated to the transition point identified by \citet{Liu2024_PopIII},
\begin{equation}
M_{\rm s,b}(\dmdtin) = 2.0 \times 10^3\,\msun \left( \frac{\dmdtin} {0.08\,\msunyr} \right)^{2.7} \ ,
\label{eq:Mstar_dMdt_Liu24}
\end{equation}
which represents the rapid growth of a supergiant protostar under very high accretion rates.

In practice, we combine these two branches and impose a conservative upper cap to mimic the lifetime and general-relativistic-instability limits,
\begin{equation}
\Mstar(\dmdtin) = \min \bigl[ \max \bigl( M_{\rm s,f}(\dmdtin), M_{\rm s,b}(\dmdtin) \bigr), 10^5\,\msun \bigr] \ .
\label{eq:Mstar_dMdt}
\end{equation}
We have verified that this simplified prescription closely follows the fiducial \citet{Liu2024_PopIII} relation over the range of ($\dmdtin$) spanned by our clouds, while being straightforward to apply to the large number of merger-tree branches analysed in this work.

Following \cite{Liu2024_PopIII}, based on the radiation-hydrodynamics results of \cite{Sugimura2023}, we adopt the conversion
\begin{equation}
  \dot{M}_{\rm in,12} = 3.7\,\dot{M}_{\rm in,7} \simeq 3.7\,\dot{M}_{\rm in,6} \ ,
\end{equation}
where $\dot{M}_{\rm in,12}$ denotes the inflow rate corresponding to \cite{Hirano2014} definition at $n_{\rm max}=10^{12}\,\cc$.
Strictly speaking, the factor $3.7$ was calibrated for the evolution of the accretion rate from the stage $n_{\rm max}=10^7$ to $10^{12}\,\cc$, whereas our cosmological simulations are terminated at $\nth=10^6\,\cc$.
We therefore use this factor as an approximate extrapolation, assuming that the additional evolution of the accretion rate between $10^6$ and $10^7\,\cc$ is small compared with the overall change from $10^6$ to $10^{12}\,\cc$.
This approximation is partly motivated by the fact that the inflow rate is evaluated at the Jeans radius, which lies outside the central $n\sim10^6\,\cc$ region and should be less sensitive to the detailed subsequent collapse of the innermost gas.

\begin{figure}
\begin{center}
\includegraphics[width=1.0\linewidth]{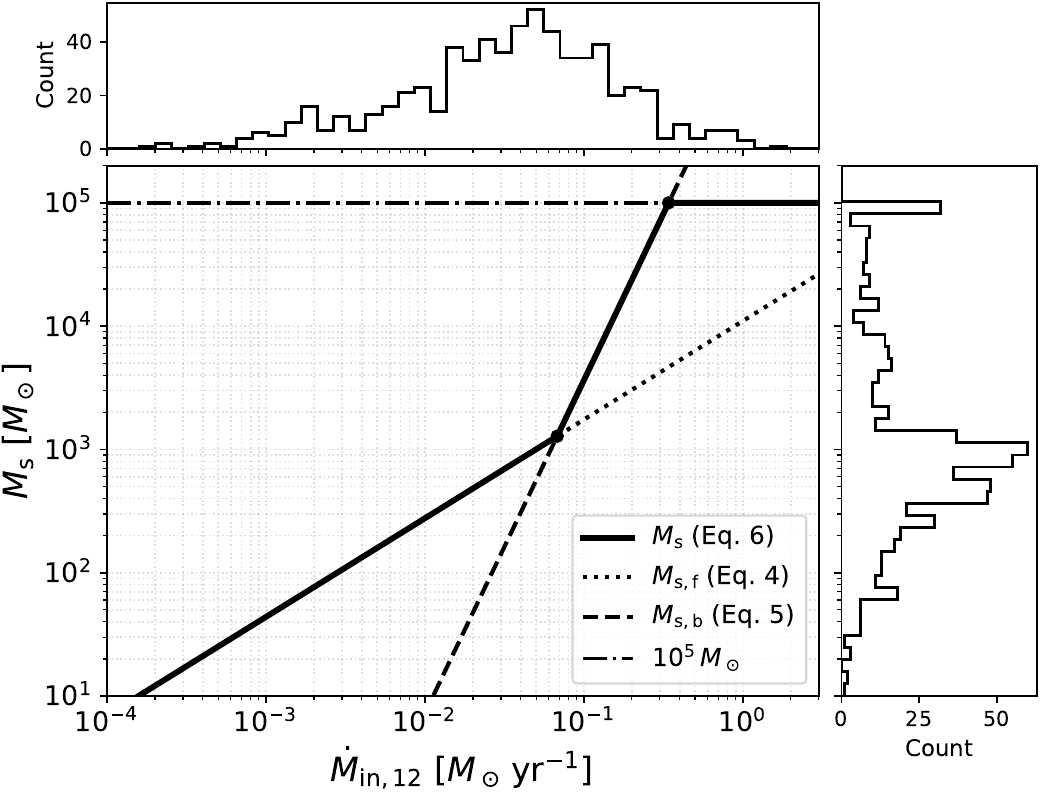}
\end{center}
\caption{
Mapping between the inflow rate and the assigned stellar mass adopted in Section~\ref{sec:methods:mapping}.
The main panel shows $M_{\rm s}$ as a function of $\dot{M}_{\rm in,12} \simeq 3.7\,\dot{M}_{\rm in,6}$.
The solid line shows the final prescription in Equation~\ref{eq:Mstar_dMdt}, while the dotted and dashed curves show the limiting relations in Equations~\ref{eq:Mstar_dMdt_Hirano14} and \ref{eq:Mstar_dMdt_Liu24}, respectively.
The dot-dashed horizontal line indicates the imposed upper cap at $10^5\,\msun$.
The top and right panels show the distributions in this study.
}
\label{fig:dMdt-Mstar_mapping}
\end{figure}

Figure~\ref{fig:dMdt-Mstar_mapping} shows the adopted mapping between $\dot{M}_{\rm in,12}$ and the assigned stellar mass $M_{\rm s}$, together with the distributions of $3.7 \dot{M}_{\rm in,6}$ and $M_{\rm s}$ in this study.

\subsubsection{Propagation of the star-formation flag}

Once a node has formed a first star, we assign a permanent ``star-formation flag'' to its merger-tree branch.
This flag is propagated to all descendant nodes.
If a progenitor has formed a first star at any earlier snapshot, all of its descendants are considered to belong to a branch that has already experienced first-star formation. 
In the merger-tree-based star-formation model, once a node forms a first star (i.e. the star-formation flag is set), we do not allow any further first-star formation in that node.
This prescription effectively mimics the suppression of subsequent first-star formation by protostellar radiative feedback, which heats and photo-evaporates the surrounding gas.
Consequently, each branch in the merger tree is allowed to trigger first-star formation at most once.
Subsequent nodes on the same branch are not permitted to form additional first stars.

However, different branches can later merge.
When two or more branches that each carry a star-formation flag merge into a single descendant node, that node inherits the first stars from all of its progenitors.
In our implementation, we therefore keep track of the number of first stars associated with each node as the sum of the first-star events that have occurred along all of its progenitor branches.
As a result, a single node can host multiple first stars.
In the merger trees analysed in this work, we find that a single node can contain up to 14 first stars.

\section{Results}
\label{sec:res}

\subsection{Overview of the halo sample}
\label{sec:res:overview}

\begin{figure}
\begin{center}
\includegraphics[width=1.0\columnwidth]{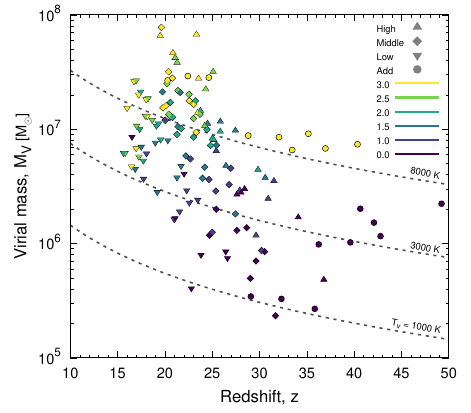}
\end{center}
\caption{
Formation redshift and virial mass of all first-star-forming haloes in our simulation suite.
The horizontal axis shows the redshift $z$ at which the gas first reaches the density threshold for cloud identification, and the vertical axis shows the corresponding virial mass $\Mvir$.
Marker shapes indicate the three halo classes: upward triangles for Class {\it High}, diamonds for Class {\it Middle}, and downward triangles for Class {\it Low}.
Filled circles labelled {\it Add} denote 18 additional runs with enhanced $\sigma_8$ that extend the sample to higher redshifts and virial masses.
Colours encode the streaming-velocity amplitude in units of the root-mean-square value, $\vsvsigma = 0$, 1, 1.5, 2, 2.5, and 3.
Dashed curves show the virial masses for three different virial temperatures, $\Tvir = 1000$, $3000$, and $8000$\,K.
}
\label{fig:z-Mvir_SV120+18}
\end{figure}

Before analysing the internal properties of clouds and first stars, we first summarise the basic characteristics of our halo sample.
Figure~\ref{fig:z-Mvir_SV120+18} shows the formation redshift and virial mass of all first-star-forming haloes, colour-coded by the streaming velocity and grouped into the three halo classes introduced in \citetalias{Hirano2025FSC2}.

The haloes cover a wide range of formation redshifts, $z \simeq 15$--$35$, and virial masses, $\Mvir \simeq 10^{6}$--$10^{8}\,\msun$.
Class {\it High} haloes populate the upper part of the diagram and typically form at $z \gtrsim 25$ with virial masses $\Mvir \gtrsim {\rm a~few}\times10^{6}\,\msun$.
Class {\it Middle} haloes occupy intermediate redshifts and virial masses, while Class {\it Low} haloes tend to form later, at $z \lesssim 20$, and include a larger fraction of lower-mass systems.

The additional 18 runs with enhanced $\sigma_8$ are shown as circular symbols labelled {\it Add}.
They extend the coverage to higher redshifts (up to $z \sim 40$--$50$), reaching virial masses of $\Mvir \sim 10^{7}\,\msun$ at these epochs, and thus provide first-star-forming haloes that are more massive and earlier formed than those in the original sample.
These additional models are particularly useful for probing the environments relevant to the most distant galaxies and AGN recently discovered by \textit{JWST}.

The streaming velocity introduces a systematic trend in the $z$-$\Mvir$ plane.
For a given class, haloes in regions with higher streaming velocities, $\vsvsigma \gtrsim 2$, tend to collapse at slightly lower redshifts and with larger virial masses than their low-streaming counterparts.
This behaviour reflects a delay in gas collapse and enhanced DM growth in high-$\vsv$ regions, consistent with previous studies of the streaming-velocity effect \citep[e.g.][]{Tseliakhovich2010, Fialkov2014}.

Overall, Figure~\ref{fig:z-Mvir_SV120+18} demonstrates that our halo sample spans a wide range of formation epochs, halo masses, and streaming velocities.
The Class {\it Low} haloes with low streaming velocities are natural analogues of the progenitors of present-day low-mass galaxies.
In contrast, Class {\it High} and Class {\it Middle} haloes provide a laboratory for exploring first-star formation in the environments of high-redshift luminous galaxies and massive BH seeds.
This is particularly true for intermediate- and high-streaming regions and for the $\sigma_8$-boosted runs.
In the following subsections, we investigate how the multiplicity and mass spectrum of first stars depend on these halo properties.

\subsection{Representative first star cluster halo}
\label{sec:res:example_I08S08V20}

\begin{figure}
\begin{center}
\includegraphics[width=1.0\linewidth]{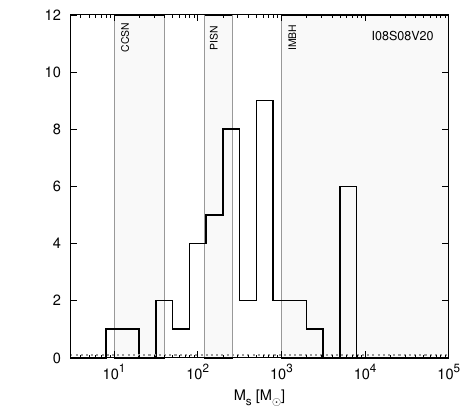}
\end{center}
\caption{
Histogram of first-star masses, $\Mstar$, in the model I08S08V20.
Hatched regions indicate the mass ranges $10 < \Mstar/\msun < 40$ (CCSN progenitors), $120 < \Mstar/\msun < 260$ (PISN progenitors), and $\Mstar/\msun > 10^{3}$ (IMBH-seed regime).
}
\label{fig:imf-vsv_I08S08V00}
\end{figure}

\begin{figure*}
\centering
    \includegraphics[width=0.49\textwidth]{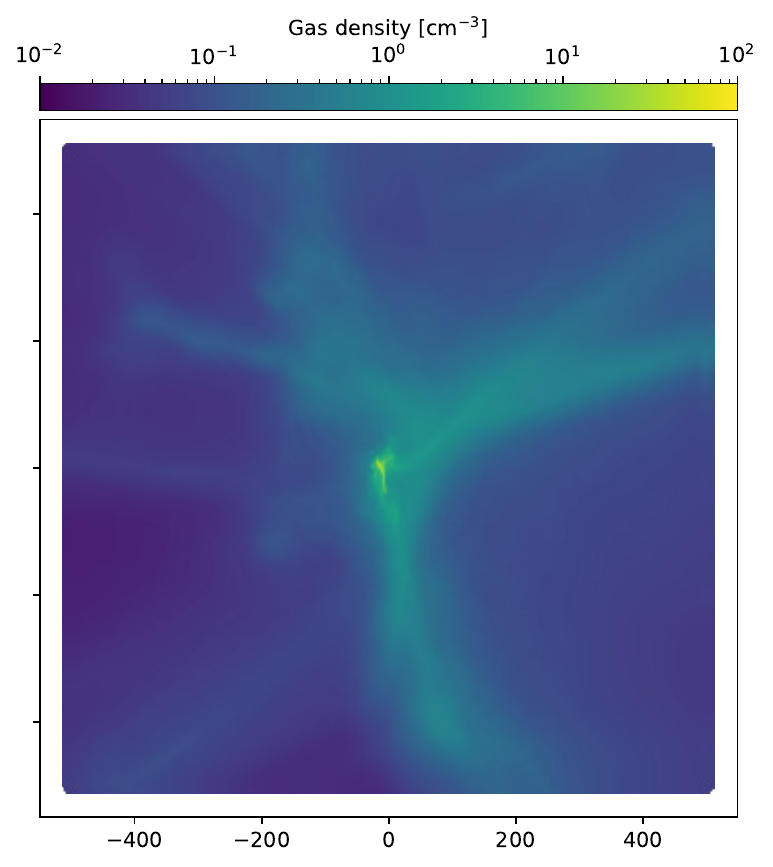}
    \includegraphics[width=0.49\textwidth]{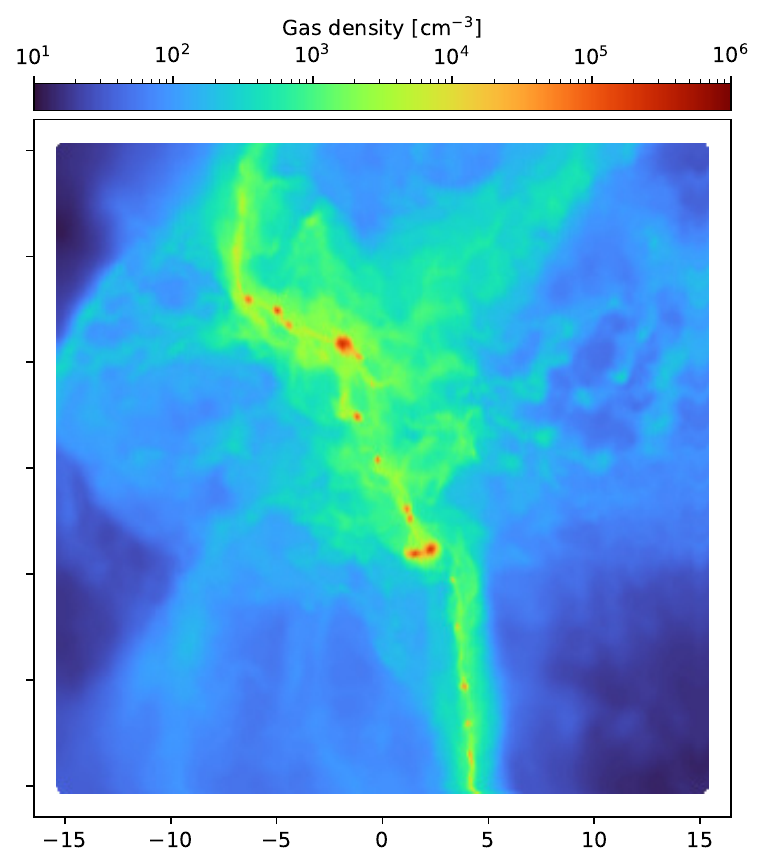}
\caption{
Gas-density structure of the representative model I08S08V20 at $\tth=2$\,Myr after the first cloud in the halo reaches the threshold density $\nth=10^6\,\cc$.
The left panel shows the projected gas density in a cube of side length 1\,kpc centred on the halo, encompassing the virial radius $\Rvir \simeq 300$\,pc.
Large-scale filaments feed the central region while the initial baryon-DM streaming velocity with $\vsvsigma=2$ is directed from left to right in this projection.
The right panel zooms in on the central 30\,pc, where a dense filament has fragmented into multiple clumps with peak densities exceeding $\nth$.
These clumps correspond to the Jeans-unstable clouds identified as nodes in the merger-tree analysis and host the formation of $\Nstar=44$ first stars in this halo over the subsequent 2\,Myr.
}
\label{fig:map2d_I08S08V20}
\end{figure*}

Before analysing the statistical properties of the halo sample, we first highlight the structure of a single FSC halo.
We select the halo I08S08V20, which forms $\Nstar = 44$ first stars over the 2\,Myr after the first cloud in the halo reaches $\nth=10^6\,\cc$.
This system belongs to the High-class haloes and experiences a relatively large initial streaming velocity, $\vsvsigma=2$, a condition under which our merger-tree analysis finds the highest cloud and star multiplicities.
Figure~\ref{fig:imf-vsv_I08S08V00} shows the resulting first-star mass spectrum in this halo, highlighting the contributions from the core-collapse supernova (CCSN), pair-instability supernovae (PISN), and intermediate-mass black hole (IMBH)-seed mass ranges.

Figure~\ref{fig:map2d_I08S08V20} shows the projected gas density around this halo at $t_{\rm th}=2$\,Myr.
On the halo scale (left panel), baryons are supplied along several large-scale filaments that converge near the halo centre, within a region comparable to the virial radius $\Rvir\simeq 300$\,pc.
The supersonic streaming motion between baryons and DM is oriented from left to right in this projection, enhancing the anisotropic inflow and promoting filamentary accretion onto the central region.

The zoomed-in view of the central 30\,pc (right panel) reveals that the main filament has fragmented into a chain of dense clumps, each reaching or exceeding the density threshold $\nth$ adopted to define Jeans-unstable clouds.
These clumps correspond to the nodes in our merger-tree analysis and host the formation of multiple first stars along the filament.
Thus, even a single halo can assemble a rich FSC through the combined effects of large-scale filamentary accretion and small-scale fragmentation, providing a concrete example of the FSC formation scenario.

To illustrate how the internal structure of star-forming clouds shapes the diversity of first-star masses, we examine in detail the representative rich FSC system in our sample, the high-$\vsv$ halo I08S08V20, which forms $\Nstar = 44$ first stars over $2$\,Myr.
For four representative stars in this halo, spanning final masses from $\Mstar \sim 10\,\msun$ up to $\sim 10^4\,\msun$, we compute spherically averaged radial profiles of gas density and temperature at their individual formation epochs.
Here, ``formation'' is defined as the time when the central density of the corresponding collapsing cloud first exceeds the threshold $\nth=10^6\,\cc$ used to construct the merger tree.

\begin{figure}
\begin{center}
\includegraphics[width=1.0\linewidth]{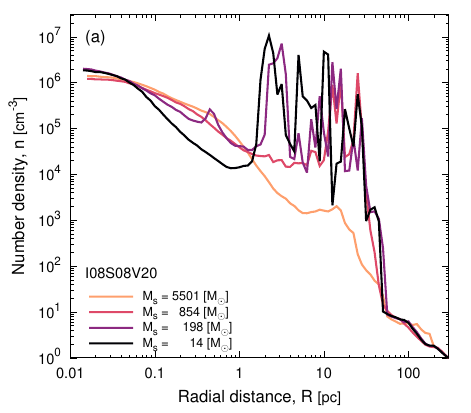}\\
\includegraphics[width=1.0\linewidth]{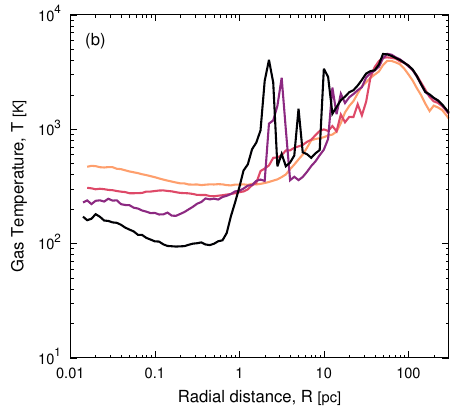}
\end{center}
\caption{
Radial gas profiles around four representative first stars in the model I08S08V20 at their formation epochs (defined by the first crossing of $\nth=10^6\,\cc$ in the collapsing cloud).
Panel (a) shows the spherically averaged number-density profiles as a function of radius from each protostar.
Panel (b) shows the corresponding gas temperature profiles.
Different coloured lines correspond to stars with masses spanning $\Mstar \simeq 10$--$10^4\,\msun$, as indicated in the legend.
}
\label{fig:Radi-Dens+Temp_I08S08V20}
\end{figure}

Figure~\ref{fig:Radi-Dens+Temp_I08S08V20} shows the resulting number-density and temperature profiles as functions of radius from each protostar.
The local density enhancement at $R \sim 2$--$10$\,pc reflects the transition from the central collapsing cloud to the surrounding dense filamentary structure (right panel of Figure~\ref{fig:map2d_I08S08V20}) because the profiles are spherically averaged around each protostar, and nearby filament/clump material appears as a bump in the radial density profile.
Lower-mass stars typically form in clouds with relatively shallow density gradients and efficient molecular cooling, reaching temperatures of a few $10^2$--$10^3$\,K at $n \sim 10^6\,\cc$.
In contrast, the most massive stars originate from clouds that remain substantially hotter during collapse, with temperatures of a few $\times10^2$\,K at comparable densities, and exhibit higher central densities at a given radius.
These differences imply systematically higher mass inflow rates onto the protostars in the hotter, more compact clouds, which our fitting formula converts into larger final stellar masses.

The comparison within a single halo highlights that the broad first-star mass spectrum is not solely a consequence of halo-to-halo variations in global properties.
It also reflects cloud-to-cloud differences in the local thermal evolution and collapse dynamics.
Even within the same large-scale environment (fixed halo mass, redshift, and streaming velocity), individual filaments and clouds can follow distinct thermodynamic paths, resulting in a wide range of accretion histories and stellar masses.
This internal diversity contributes to the substantial halo-to-halo scatter in the per-halo IMFs discussed in Section~\ref{sec:res:imf}, and must be taken into account when modelling first-star formation with simplified sub-grid prescriptions.

\subsection{Multiplicity of clouds and first stars}
\label{sec:res:Vsv-multiplicity}

Having examined in Section~\ref{sec:res:example_I08S08V20} how individual clouds in a representative halo produce a wide range of first-star masses, we now turn to the full halo sample and quantify how cloud and stellar multiplicities depend on the halo environment.

\begin{figure*}
\centering
    \includegraphics[width=0.49\textwidth]{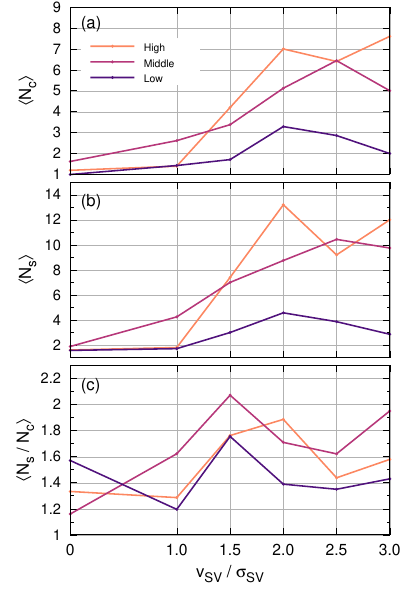}
    \includegraphics[width=0.49\textwidth]{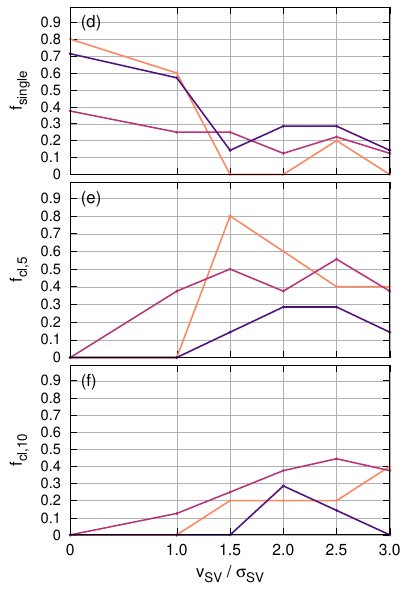}
\caption{
Panels (a)-(c) show the mean numbers of clouds and first stars per halo as a function of the streaming velocity, $\vsvsigma$.
Lines indicate the three halo classes: {\it High} (orange), {\it Middle} (red), and {\it Low} (purple).
We plot the mean number of Jeans-unstable clouds $\langle \Ncore \rangle$ in panel (a), the mean number of first stars $\langle \Nstar \rangle$ in panel (b), and the mean multiplicity per cloud $\langle \Nstar/\Ncore \rangle$ in panel (c).
Panels (d)-(f) show the corresponding fractions of haloes hosting different numbers of first stars.
We plot the fraction of single-star haloes $f_{\rm single}$ in panel (d), the fraction of haloes with at least five first stars $f_{\rm cl,5}$ in panel (e), and the fraction with at least ten first stars $f_{\rm cl,10}$ in panel (f).
}
\label{fig:Vsv-Ncloud-Nstar-rate}
\end{figure*}

Figure~\ref{fig:Vsv-Ncloud-Nstar-rate} summarizes how the multiplicity of dense clouds and first stars depends on the streaming velocity for the three halo classes.
The left panels show the mean numbers of clouds and first stars per halo, $\langle \Ncore\rangle$ and $\langle \Nstar\rangle$, and the mean stellar multiplicity per cloud, $\langle \Nstar\rangle/\langle \Ncore\rangle$.
The right panels show the fractions of haloes hosting at least one, five, or ten first stars: $f_{\rm single}$, $f_{\rm cl,5}$, and $f_{\rm cl,10}$.

Across all Classes, the mean number of clouds per halo increases with the streaming velocity.
At $\vsvsigma \lesssim 1$, typical haloes form only one or two Jeans-unstable clouds, $\langle \Ncore\rangle \simeq 1$--2.
At $\vsvsigma \gtrsim 1.5$, this rises to $\langle \Ncore \rangle \sim 4$--$7$, depending on the Class.
The mean number of first stars per halo, $\langle \Nstar \rangle$, follows the same trend.
It increases from $\simeq 2$--$3$ at low streaming velocities to $\sim 8$--$11$ for Class {\it High} and Class {\it Middle} haloes at $\vsvsigma \gtrsim 2$.
Class {\it Low} haloes remain less fragmenting overall, typically with $\langle \Ncore \rangle \lesssim 3$ and $\langle \Nstar\rangle \lesssim 4$ even in the highest-$\vsv$ bins.

The mean stellar multiplicity per cloud stays close to unity across all Classes.
We find $\langle \Nstar/\Ncore \rangle \simeq 1.2$--$2.0$ in most groups, with only a mild dependence on $\vsvsigma$.
Thus, the primary role of streaming is to increase the number of distinct self-gravitating clouds per halo, rather than to strongly increase the cloud-merger frequency and thus the number of first stars per cloud after merging.

The fractions of haloes with different levels of first-star multiplicity reveal a complementary picture.
In the low-streaming regime, single-star haloes are common, with $f_{\rm single} \sim 0.6$--$0.7$ for the low-$\vsv$ groups in Class {\it High} and Class {\it Low}, and $f_{\rm single} \sim 0.3$ for Class {\it Middle}.
In this regime, haloes with rich FSCs are rare, and the fractions $f_{\rm cl,5}$ and $f_{\rm cl,10}$ remain close to zero.
As the streaming velocity increases, $f_{\rm single}$ rapidly declines, dropping below $\sim 0.2$ at $\vsvsigma \gtrsim 1.5$ for all Classes.

Conversely, the cluster fractions $f_{\rm cl,5}$ and $f_{\rm cl,10}$ grow with the streaming velocity.
In particular, Class {\it High} haloes with intermediate and high streaming velocities frequently host FSCs with $\Nstar \ge 5$ and even $\Nstar \ge 10$, with $f_{\rm cl,5} \gtrsim 0.4$ and $f_{\rm cl,10} \gtrsim 0.2$ in these bins.
Class {\it Middle} haloes show a similar but slightly weaker behaviour, while Class {\it Low} haloes rarely reach $\Nstar \ge 10$ but still attain $f_{\rm cl,5} \sim 0.3$--$0.4$ at $\vsvsigma \gtrsim 2$.
These trends confirm that supersonic streaming promotes the formation of multiple clouds and, as a result, favours the emergence of FSCs rather than isolated first stars.
We note that the binned trends in Figure~\ref{fig:Vsv-Ncloud-Nstar-rate} are based on group averages, and thus small non-monotonic variations between adjacent $\vsvsigma$ bins can occur due to finite sample sizes and halo-to-halo scatter.

\begin{figure}
\begin{center}
\includegraphics[width=1.0\linewidth]{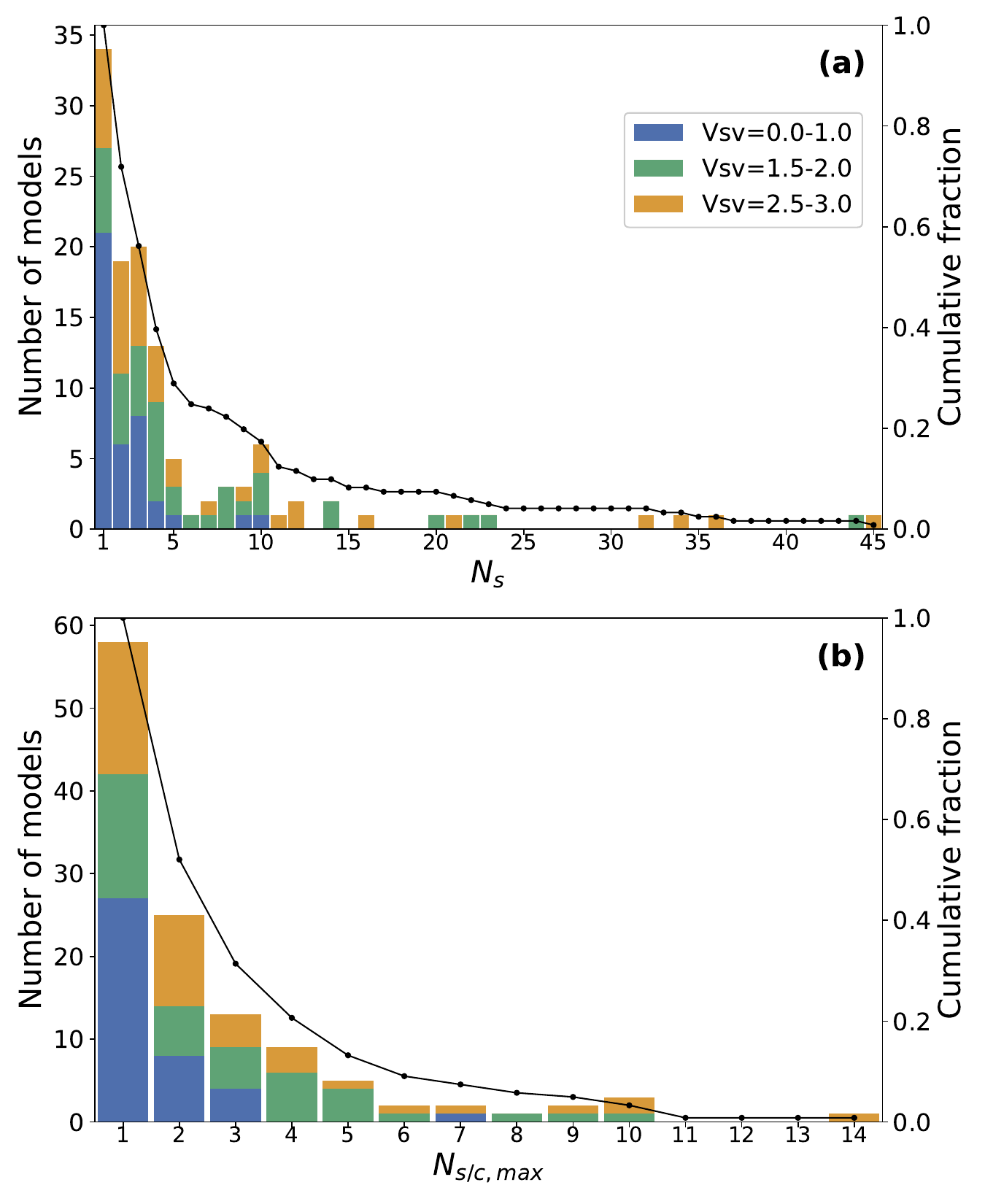}
\end{center}
\caption{
Distributions of the number of first stars formed in each halo and within individual clouds.
Panel~(a) shows the histogram of the total number of first stars per halo, $\Nstar$, stacked by streaming velocity in three ranges: $\vsvsigma=0$--$1$ (blue), $1.5$--$2$ (green), and $2.5$--$3$ (orange).
Panel~(b) shows the histogram of the maximum number of stars in a single cloud within each halo, $\Nstarcloud$.
The black lines indicate the cumulative fraction of models.
}
\label{fig:hist_Ns_Nsc}
\end{figure}

Figure~\ref{fig:hist_Ns_Nsc} shows distributions of the number of first stars formed in each halo and within individual clouds.
Figure~\ref{fig:hist_Ns_Nsc}(a) shows the histogram of the total number of first stars per halo, $\Nstar$, stacked by streaming velocity in three ranges: $\vsvsigma=0$--$1$ (blue), $1.5$--$2$ (green), and $2.5$--$3$ (orange).
The black line with dots indicates the cumulative fraction of haloes as a function of $\Nstar$.
Most haloes form only a few first stars ($\Nstar \la 3$), while a small number of systems host up to $\Nstar\sim 40$.
Figure~\ref{fig:hist_Ns_Nsc}(b) shows the histogram of the maximum number of stars in a single cloud within each halo, $\Nstarcloud$.
In the majority of haloes, the most populated cloud hosts only one or two first stars, but a non-negligible tail of systems reaches $\Nstarcloud \ga 5$, corresponding to compact FSCs.
High-velocity models tend to populate the multi-star and cluster tail more frequently than low-velocity ones.

To verify the same overall tendencies with reduced halo-to-halo scatter, we also combine the models into three broader streaming-velocity regimes, $\vsvsigma=0$--$1$, $1.5$--$2$, and $2.5$--$3$, and summarize the corresponding group-averaged multiplicity statistics in Table~\ref{table:group_class-vs_mean}.
This table should be read as a coarse-bin summary of the trends already shown in Figure~\ref{fig:Vsv-Ncloud-Nstar-rate}, rather than as an independent primary result.
In particular, it confirms that the low-streaming groups are dominated by single-star haloes, whereas the intermediate- and high-streaming groups preferentially host FSCs with large values of $\Ncore$, $\Nstar$, $f_{\rm cl,5}$, and $f_{\rm cl,10}$.

A more detailed view of the full mass functions, their dependence on halo environment, and the halo-to-halo scatter around the mean trends is presented in Section~\ref{sec:res:imf}.

\subsection{Environmental dependence of stellar mass functions}
\label{sec:res:imf}

\begin{figure}
\begin{center}
\includegraphics[width=1.0\columnwidth]{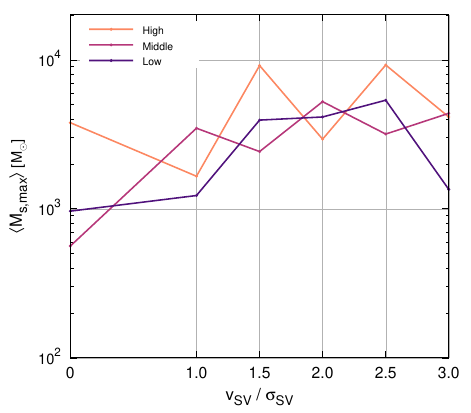}
\end{center}
\caption{
Mean of the maximum stellar mass, $\langle \Mstarmax \rangle$, as a function of the streaming velocity, $\vsvsigma$, for Class {\it High} (orange), {\it Middle} (red), and {\it Low} (purple) haloes.
}
\label{fig:Vsv-Mstarmax}
\end{figure}

\begin{figure}
\begin{center}
\includegraphics[width=1.0\columnwidth]{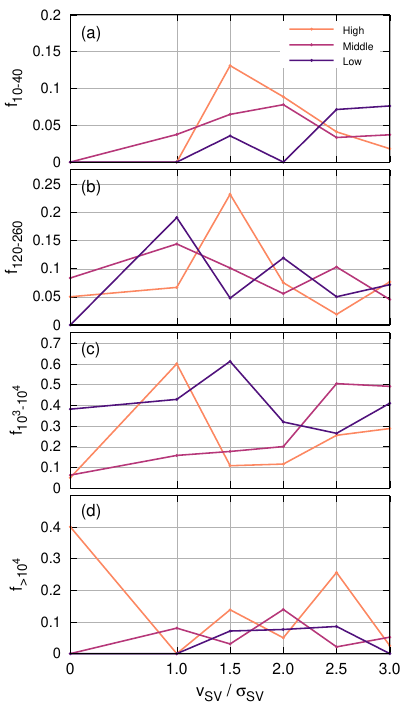}
\end{center}
\caption{
Fraction of stellar mass as a function of the streaming velocity, $\vsvsigma$, for Class {\it High} (orange), {\it Middle} (red), and {\it Low} (purple) haloes.
Panels (a)-(d) show the mean fractions of stars with 10--40, 120--260, $10^3$--$10^4$, and $\ge 10^4\,\msun$, respectively.
}
\label{fig:Vsv-fmassbin}
\end{figure}

\begin{figure*}
\begin{center}
\includegraphics[width=1.0\linewidth]{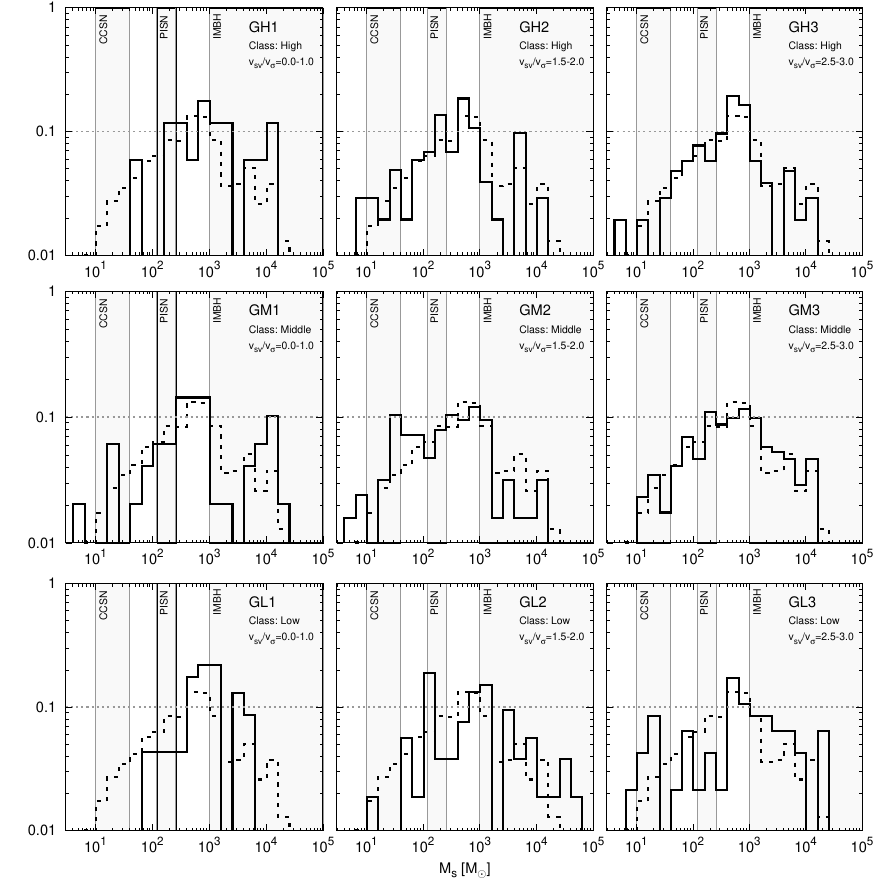}
\end{center}
\caption{
Group-averaged IMFs of first stars.
Rows show the three halo classes (H/M/L) and columns show the streaming-velocity bins (1/2/3).
Solid histograms show group IMFs (averages of normalised per-halo histograms) and dashed histograms show the all-halo average.
Vertical grey lines mark $\Mstar/\msun=10$--40, 120--260, and $\ge 10^3$.
}
\label{fig:imf-group_3times3}
\end{figure*}

We now examine how the stellar-mass spectrum itself depends on the formation environment.
We first summarise the main trends of the scalar indicators of the first-star mass spectrum shown in Figures~\ref{fig:Vsv-Mstarmax} and \ref{fig:Vsv-fmassbin}.
Across all halo Classes and streaming-velocity ranges, the typical maximum stellar mass per halo is of order $10^{3}\,\msun$.
Therefore, very massive first stars are a common outcome in our sample.
Low-$\vsv$ haloes tend to have somewhat lower maximum masses.
At higher $\vsv$, $\langle \Mstarmax \rangle$ is typically a few $\times 10^{3}\,\msun$, but varies non-monotonically between bins and can reach $\sim\!10^{4}\,\msun$ in some environments.

The mass-bin fractions highlight three particularly important regimes.
First, the fraction of stars in the $10$--$40\,\msun$ range, $f_{10-40}$, is small in all environments, typically at the level of a few per cent and generally below $\sim\!0.1$, occasionally reaching $\sim\!0.1$--$0.15$.
Within our halo sample, first-star formation is therefore strongly biased toward either lower masses below $10\,\msun$ or higher masses above $40\,\msun$, with only a minor contribution from the $10$--$40\,\msun$ interval that is most relevant for enriching long-lived second-generation stars in the Milky Way.

Second, the PISN progenitor range, $120 \le \Mstar/\msun < 260$, carries a substantial fraction $f_{120-260} \sim 0.1$--$0.3$ in many groups, especially for Class {\it High} and Class {\it Middle} haloes.
Even Class {\it Low} haloes maintain non-negligible values $f_{120-260} \sim 0.1$--$0.2$, implying that PISN-capable first stars are not rare in the environments we consider.

Third, the very high-mass tail, $\Mstar \ge 10^3\,\msun$, is also prominent.
In particular, high-mass haloes with moderate or large streaming velocities have $f_{>10^3}$ (computed as $f_{10^3\text{--}10^4}+f_{>10^4}$) of order $\sim 0.2$--$0.4$, and even some lower-Class systems retain a tail of very massive stars.
These scalar indicators show that cluster-forming, high-$\vsv$ environments generically host very massive and potentially supermassive first stars, while simultaneously producing only a small fraction of stars in the $10$--$40\,\msun$ range.

Table~\ref{table:group_class-vs_imf} summarizes the corresponding group-averaged mass-bin statistics for the same three broader streaming-velocity regimes.
It confirms that the $10$--$40\,\msun$ fraction remains small across all environments, while the PISN range and the very-high-mass tail retain non-negligible contributions in many intermediate- and high-streaming groups.

We now turn to the full first-star mass functions and discuss both their systematic dependence on halo environment and their halo-to-halo scatter.
Figure~\ref{fig:imf-group_3times3} shows the binned first-star mass functions for haloes grouped by Class (High, Middle, Low) and streaming-velocity range ($\vsvsigma = 0$--$1$, $1.5$--$2$, $2.5$--$3$).
Within each panel, we stack the IMFs of all haloes in the corresponding group and normalise them by the total number of stars.
The resulting histograms represent the probability distribution of stellar masses in that environment.

Several systematic features are apparent.
First, the characteristic stellar mass decreases from the upper left to the lower right of the grid.
High-Class haloes, which typically collapse earlier and in larger virial masses, exhibit IMFs that are strongly top-heavy, with a large fraction of stars at $\Mstar \gtrsim 10^3\,\msun$ and very few objects below $40\,\msun$.
In contrast, Low-Class haloes, forming later and including a larger fraction of lower-mass hosts, show a broader distribution extending down to $\Mstar \sim 10$--$40\,\msun$ and a less prominent very-high-mass tail.
Second, at fixed Class, increasing streaming velocity tends to shift probability toward higher stellar masses and to enhance the high-mass tail, broadly consistent with the trends already seen in the scalar indicators (Figures~\ref{fig:Vsv-Mstarmax} and \ref{fig:Vsv-fmassbin}).
Finally, across all groups, the fraction of stars in the $10$--$40\,\msun$ range remains relatively small, while the bins corresponding to very massive stars, $\Mstar \gtrsim 120\,\msun$ and especially $\Mstar \gtrsim 10^3\,\msun$, retain non-negligible probabilities in many environments.

Thus, both the scalar indicators and the full stacked mass functions point to the same overall conclusion: first star formation in high-$\vsv$ environments is not only more clustered, but also more strongly weighted toward the very-high-mass end.

\section{Discussion}
\label{sec:dis}

\subsection{Quantifying halo-to-halo IMF variations}
\label{sec:tv_distance}

The universality of the stellar IMF has long been tested in nearby star-forming systems using population-statistical comparisons \citep[e.g.][]{Kroupa2001,Dib2017,Singh-Bal2025}.
Here we adopt a non-parametric analogue for Pop~III star formation by comparing discretised per-halo stellar-mass distributions to reference templates.

For each halo $h$, we define a binned stellar-mass PDF $p_i^{(h)}$, where $p_i^{(h)}$ is the fraction of stars in the $i$-th mass bin.
We construct two reference IMFs: the global mean $p_i^{\rm (all)}$ (averaged over all haloes) and the group mean $p_i^{\rm (grp)}$ (averaged over haloes in the same Class and streaming-velocity range as $h$).
We quantify the deviation from each reference using the total-variation distance,
\begin{equation}
D_{\rm TV}(p,q) = \frac{1}{2} \sum_i \left| p_i - q_i \right| \ ,
\end{equation}
and compute
\begin{eqnarray}
D_{\rm all}(h) = D_{\rm TV}\!\left( p^{(h)}, p^{\rm (all)} \right) \ ,\\
D_{\rm grp}(h) = D_{\rm TV}\!\left( p^{(h)}, p^{\rm (grp)} \right) \ .
\end{eqnarray}

\begin{figure*}
\begin{center}
\includegraphics[width=1.0\linewidth]{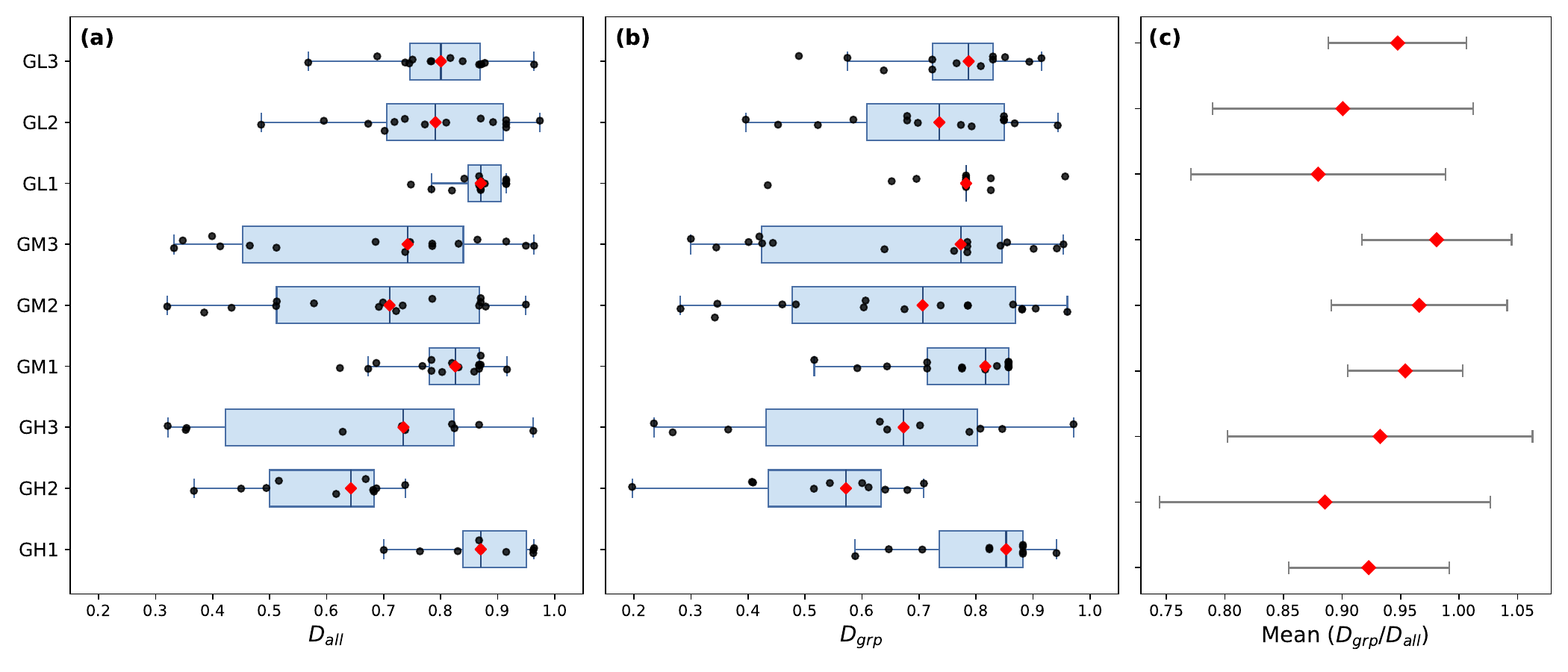}
\end{center}
\caption{
Halo-to-halo variations of the first-star mass functions.
For each halo, we compute the total-variation distance between the per-halo stellar-mass distribution and (a) the all-halo averaged IMF, $D_{\rm all}$, and (b) the group-averaged IMF, $D_{\rm grp}$.
Groups are defined by the halo class and streaming-velocity bin (GH1--GL3; Table~2).
Black points show individual haloes, blue boxes show interquartile ranges, and red diamonds show means.
For GL1 in panel (b), the interquartile range is exactly zero ($Q1=Q3$), so the box collapses and may appear invisible (see Appendix~\ref{app:binwidth_test}).
Panel (c) shows the group-mean ratio $\langle D_{\rm grp}/D_{\rm all}\rangle$.
}
\label{fig:imf-dist}
\end{figure*}

Figure~\ref{fig:imf-dist} shows that $D_{\rm all}$ is typically $\sim\!0.5$--$0.9$ (with values extending up to $\sim\!1$).
In all Classes and streaming-velocity bins, $D_{\rm grp}$ is systematically smaller than $D_{\rm all}$, indicating that environment-conditioned templates provide a better statistical description of the per-halo IMFs, although substantial scatter remains.

This scatter has two components.
First, it reflects small-number sampling: most haloes form only a few first stars (median $\Nstar \sim 2$; Figure~\ref{fig:hist_Ns_Nsc}), so the per-halo PDF can deviate strongly from any underlying template.
Second, it is physical: the systematic differences among group-averaged IMFs (Figure~\ref{fig:imf-group_3times3}) imply that the underlying Pop~III IMF depends on environment, and that early-forming haloes and, in many cases, higher-$\vsv$ bins exhibit a more pronounced high-mass tail.
Both effects contribute to the large values of $D_{\rm all}$: sampling noise is amplified when $\Nstar$ is small, and the mismatch is further increased when a single universal template is assumed across distinct environments.

A practical implication is that a single universal Pop~III IMF is unlikely to be an adequate approximation across the full range of environments probed here.
Instead, the group-averaged IMFs can be treated as environment-dependent templates, $\phi(\Mstar|{\cal E})$, and per-halo outcomes as stochastic realisations with small $\Nstar$.
This framework helps interpret both chemical-abundance constraints in metal-poor stars (which often favour enrichment by a single or a few progenitors when $\Nstar$ is small) and the high-mass tail relevant for massive remnants.

\subsection{Connections to observed galaxies and BHs}
\label{sec:dis:obs}

\begin{figure*}
\begin{center}
\includegraphics[width=1.0\linewidth]{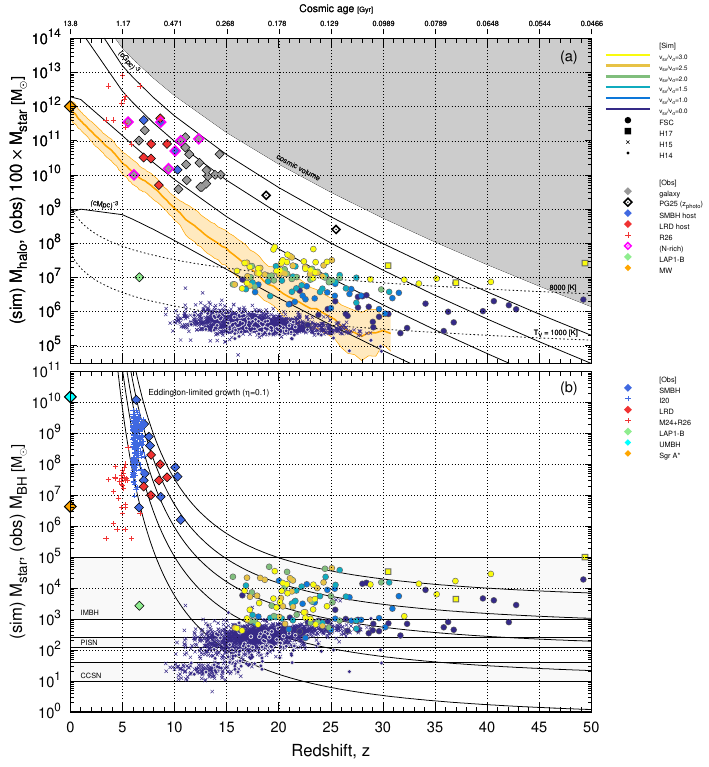}
\end{center}
\caption{
Redshift dependence of halo and stellar (or BH) masses compared with observations.
The upper horizontal axis shows the corresponding cosmic age.
Circles indicate our 138 simulation haloes, with colours encoding $\vsvsigma = 0$, 1, 1.5, 2, 2.5, and 3.
Squares, `$\times$' markers, and small filled dots denote simulation results from our previous work; diamonds and `$+$' markers denote observational data points (see Appendix~B for dataset definitions and references).
Panel (a): simulation points show virial halo masses, while observational points are plotted at $M_{\rm halo}=100\,M_{\ast}$ for a fiducial stellar-to-halo mass ratio of $\sim 0.01$.
Solid curves show halo masses corresponding to one object per $(1\,{\rm cMpc})^3$, $(10\,{\rm cMpc})^3$, $(100\,{\rm cMpc})^3$, and $(1\,{\rm cGpc})^3$, estimated from the Press-Schechter halo mass function.
Dashed curves indicate the virial masses for fixed virial temperatures, $\Tvir = 1000$\,K and 8000\,K.
The orange hatched region shows the range (minimum to maximum) and the mean halo-mass evolution of MW-like progenitors derived from merger trees in a large-volume cosmological simulation \citep[Phi-4096;][]{IshiyamaHirano2025}.
Panel (b): simulation points show the maximum Pop~III stellar mass in each halo, $\Mstarmax$, while observational points show central BH masses (except for LAP1-B, for which the inferred total Pop~III stellar mass is shown).
Solid curves show Eddington-limited growth tracks with radiative efficiency $\eta=0.1$, chosen to pass through $(z, M_{\rm BH}/\msun)=(40, 10^{4})$, $(30, 10^{3.5})$, $(25, 10^{3})$, $(20, 10^{2.5})$, and $(15, 10^{2})$.
Horizontal lines mark characteristic mass scales for CCSN progenitors, PISN progenitors, and the IMBH-seed regime.
}
\label{fig:z-Mhalo-Mbh}
\end{figure*}

Figure~\ref{fig:z-Mhalo-Mbh} compares our simulated haloes and their most massive first stars with representative observed galaxies and BHs in the redshift--mass plane.
Panel (a) shows virial halo masses, while panel (b) compares the maximum Pop~III stellar mass in each halo to inferred BH masses, assuming a remnant (seed) mass comparable to $\Mstarmax$, together with illustrative Eddington-limited growth tracks \citep[e.g.][]{Yuan2014}.

In panel (a), our full model sample populates $\Mvir\sim10^{6}$--$10^{8}\,\msun$ over $z\simeq50$--$15$, whereas observed galaxies at $z\sim5$--$15$ inhabit much more massive haloes, $\Mvir\sim10^{9}$--$10^{12}\,\msun$ \citep[plotted, for illustration, at $M_{\rm halo}=100\,M_{\ast}$ for a fiducial stellar-to-halo mass ratio of $\sim0.01$][]{Wechsler2018}.
Galaxies at $z\gtrsim10$ therefore lie above the mass scale directly resolved here, but in a hierarchical picture, they are assembled from many progenitors comparable to our Class {\it High/Middle} haloes at $z\gtrsim20$.
In our simulations, progenitors with large streaming velocities preferentially form FSCs and very massive stars, providing an efficient channel for early enrichment and massive remnants in the progenitor population.
At intermediate redshifts ($z\simeq5$--$10$), typical halo masses are $\Mvir\sim10^{9}$--$10^{10}\,\msun$, consistent with descendants of our Class {\it Middle/Low} progenitors formed at $z\sim20$--$15$, where we predict small multiplicities and a substantial contribution from $\Mstar \lesssim 40\,\msun$ stars.

The 18 additional $\sigma_8$-enhanced models (Section~2.1) extend the coverage to rarer, earlier-forming haloes at very high redshift.
In panel (a), these models populate the regime traced by the Press-Schechter rarity curves \citep{PressSchechter1974, ShethTormen1999}, which is relevant to progenitors of extremely rare, very high-$z$ compact galaxies \citep[e.g.][]{Perez-Gonzalez2025}.
They include cases where even $\vsvsigma = 0$ yields $\Mvir\sim10^{6}\,\msun$ at $z \sim 40$, as well as cases where $\vsvsigma = 3$ delays collapse into the atomic-cooling regime with $\Tvir \gtrsim 8000$\,K.
In panel (b), the same haloes often host very massive first stars, $\Mstarmax\sim10^{4}\,\msun$, implying seed masses that can reach the observed $(z, M_{\rm BH})$ space of high-$z$ AGN/LRDs under sustained Eddington-limited growth.

More generally, if the most massive first stars leave remnants with $M_{\rm BH}\sim\Mstarmax$, then our cluster-forming haloes with $\Mstarmax\sim10^{3}$--$10^{4}\,\msun$ provide a range of heavy-seed masses  \citep[e.g.][]{Inayoshi2020}.
The Eddington tracks indicate that seeds of $\sim10^{3}$--$10^{4}\,\msun$ formed at $z \sim 30$--20 can, in principle, grow into the BHs observed at $z\sim10$ and $z\sim7$.
By contrast, lower-mass seeds of $\sim10^{2}$--$10^{3}\,\msun$ tend to remain in the intermediate-mass regime and may map more naturally onto less extreme AGN/LRDs or local systems such as Sgr~A$^\ast$.

Overall, Figure~\ref{fig:z-Mhalo-Mbh} organises our results in terms of a small set of control parameters (redshift, halo mass, and streaming velocity).
It also provides a compact bridge between Pop~III star formation in minihalo progenitors and observed galaxies and BHs.
Our results also provide inputs for heavy-seed and gravitational-wave population models.
Pop~III clusters that efficiently form intermediate-mass remnants and binaries can contribute to intermediate-mass binary BH mergers detectable by next-generation space missions \citep[e.g. \textit{LISA} and \textit{TianQin};][]{Amaro-Seoane2023_LISA, Wang2025}.
These connections remain limited by the finite dynamical range of our simulations and by uncertainties in the accretion-rate-to-mass mapping, but the per-halo IMFs and their $(z, \Mvir, \vsv)$ dependence provide useful inputs for semi-analytic heavy-seed population models \citep[e.g.][]{Caceres-Burgos2025}.

\section{Conclusions}
\label{sec:con}

We used 138 cosmological zoom-in hydrodynamics simulations from the FSC project to test the assumption of a universal Pop~III IMF in a systematic, cosmological setting.
We constructed a post-processed dense-cloud (node) merger tree and assigned stellar masses by mapping the radial gas accretion-rate profile to a final stellar mass (a post-processed prescription that does not explicitly follow radiative feedback or subsequent fragmentation).
This yields physically motivated per-halo first-star IMFs without imposing an a priori IMF.
The resulting prescription is straightforward to incorporate into semi-numerical and galaxy-formation models.
This prescription can be tested against upcoming Cosmic Dawn/EoR probes \citep[including 21-cm measurements; e.g.][]{Gessey-Jones2025}.

Our central result is that Pop~III IMFs are not universal at the halo level.
Both the cloud multiplicity and the prominence of the high-mass tail vary strongly with environment (here characterized by $(z, \Mvir, \vsv)$).
Low-mass, low-$\vsv$ haloes typically form only one or a few first stars, whereas massive haloes in high-$\vsv$ regions host rich FSCs and frequently produce very massive ($\gtrsim 10^{3}$--$10^{4}\,\msun$) first stars.
Even at fixed $(z, \Mvir, \vsv)$, the per-halo IMF exhibits substantial halo-to-halo scatter that can be understood as the combination of (i) a statistical component driven by small numbers of star-forming events and (ii) a physical component corresponding to an environment-dependent shift in the baseline mass scale.

Placing our FSC seeds on the redshift--mass plane, we find that very massive first stars born in high-$\vsv$, high-$\Mvir$ haloes naturally populate the regime required to seed the BHs powering high-redshift quasars and LRDs \citep[e.g.][]{Rusakov2026}.
Conversely, weaker-streaming environments that form only a few first stars are more consistent with chemical signatures of one or a few enrichment events in metal-poor stars \citep[e.g.][]{Karlsson2013}, highlighting that a single, universal Pop~III IMF can be systematically misleading.
In rare cases, the same conditions that favour rich FSC formation can yield multiple very massive BH remnants and thus a massive BH pair, potentially relevant to the emerging discussion of dual compact high-$z$ systems \citep[e.g.][]{Yue2021,Tanaka2024,Yanagisawa2026}.

Beyond these implications, our analysis provides a practical, physics-based prescription for per-halo Pop~III IMFs.
The prescription can be implemented as a sub-grid model in galaxy-formation calculations by conditioning the IMF on $(z, \Mvir, \vsv)$ and cloud-scale inflow properties, while retaining intrinsic halo-to-halo diversity.
The approach is limited by the finite dynamical range of the current simulations and by uncertainties in the accretion-rate-to-mass mapping.
Future work will follow collapse to higher densities and extend the evolution to later times with radiative feedback.
These extensions will allow us to couple the resulting seed populations to BH growth models.
Such extensions will enable direct statistical comparisons with the rapidly expanding \textit{JWST}/\textit{ALMA} census and with chemical constraints from extremely metal-poor stars \citep[e.g.][]{FrebelMorris2015}.

\section*{Acknowledgements}

We thank Sunmyon Chon, Daisuke Toyouchi, Naoki Yoshida, and Kazuyuki Omukai for fruitful discussions, Hyunbae Park for discussing the BTD approximation in the cosmological initial-condition setting, and Tomoaki Ishiyama for constructing the merger-tree data.
We also thank our anonymous referee for constructive comments on this manuscript.
Numerical computations were carried out on Cray XC50 and XD2000 at CfCA in the National Astronomical Observatory of Japan and Yukawa-21 at YITP in Kyoto University.
Numerical analyses were carried out on the analysis servers at CfCA in the National Astronomical Observatory of Japan.
This work was supported by JSPS KAKENHI Grant Numbers JP21K13960 and JP21H01123 / JP23K20864.
Part of this work was supported by the NAOJ Research Coordination Committee, NINS (NAOJ-RCC-2502-0202), and Fukui Prefectural University.

\section*{Data Availability}

Additional data from this study are available from the corresponding author on reasonable request.


\bibliographystyle{mnras}
\bibliography{ms}

\appendix

\section{Results of 138 models}
\label{app:table_results_each}

\begin{table*}
\centering
\caption{Results of 120 models.}
\label{table:120models}
\begin{tabular}{lccccccccc}
\hline
Model &
$z$ &
$\Mvir$ &
$\Ncore$ &
$\Nstar$ &
$\Nstar/\Ncore$ &
$\Nstarcloud$ &
$\Mstartot$ &
$\Mstarmax$ \\
&
&
($\msun$) &
&
&
&
&
($\msun$) &
($\msun$) \\
\hline
I01S08V00             & 36.80 & $3.723\times10^5$ &  1 &  1 & 1.0 &  1 &   452 &   452 \\
I01S08V10$^{\dagger}$ & 30.91 & $1.920\times10^6$ &  1 &  1 & 1.0 &  1 &  2174 &  2174 \\
I01S08V15$^{\dagger}$ & 27.45 & $7.856\times10^6$ &  4 &  4 & 1.0 &  1 &  1238 &   891 \\
I01S08V20             & 25.84 & $1.794\times10^7$ &  6 &  8 & 1.3 &  3 &  2007 &   891 \\
I01S08V25             & 25.10 & $2.419\times10^7$ &  1 &  4 & 4.0 &  4 & 51820 & 44150 \\
I01S08V30             & 23.33 & $6.468\times10^7$ &  7 & 16 & 2.3 &  9 & 35510 & 12870 \\
I02S08V00             & 34.09 & $1.319\times10^6$ &  2 &  4 & 2.0 &  3 &  9289 &  8770 \\
I02S08V10             & 31.53 & $2.855\times10^6$ &  2 &  3 & 1.5 &  2 &  8046 &  5519 \\
I02S08V15             & 30.60 & $3.576\times10^6$ &  2 &  6 & 3.0 &  5 & 16230 & 15220 \\
I02S08V20             & 29.84 & $4.163\times10^6$ &  2 &  4 & 2.0 &  2 & 17030 & 15390 \\
I02S08V25             & 21.28 & $3.670\times10^7$ &  6 &  7 & 1.2 &  2 & 11230 &  7833 \\
I02S08V30             & 19.50 & $4.732\times10^7$ &  2 &  3 & 1.5 &  2 &  6659 &  4623 \\
I03S08V00             & 31.68 & $2.195\times10^5$ &  1 &  1 & 1.0 &  1 &   294 &   294 \\
I03S08V10             & 20.03 & $1.169\times10^7$ &  4 & 10 & 2.5 &  7 & 40940 & 17280 \\
I03S08V15$^{\dagger}$ & 21.37 & $9.943\times10^6$ &  1 &  1 & 1.0 &  1 &   911 &   911 \\
I03S08V20             & 20.81 & $1.229\times10^7$ & 11 & 23 & 2.1 &  9 & 46920 & 33380 \\
I03S08V25             & 20.41 & $1.341\times10^7$ &  6 &  9 & 1.5 &  3 &  6620 &  2580 \\
I03S08V30             & 20.13 & $1.531\times10^7$ &  3 &  3 & 1.0 &  1 &  4360 &  3521 \\
I04S08V00             & 30.53 & $7.717\times10^5$ &  3 &  3 & 1.0 &  1 &   887 &   728 \\
I04S08V10             & 30.22 & $7.972\times10^5$ &  2 &  2 & 1.0 &  1 &   800 &   544 \\
I04S08V15$^{\dagger}$ & 27.45 & $1.647\times10^6$ &  5 &  7 & 1.4 &  3 &  1588 &   489 \\
I04S08V20             & 24.82 & $6.800\times10^6$ &  1 &  1 & 1.0 &  1 &   532 &   532 \\
I04S08V25             & 23.02 & $1.183\times10^7$ &  2 &  2 & 1.0 &  1 &  5608 &  4266 \\
I04S08V30             & 23.31 & $8.841\times10^6$ &  1 &  1 & 1.0 &  1 &  2045 &  2045 \\
I05S08V00             & 29.63 & $6.011\times10^5$ &  1 &  1 & 1.0 &  1 &   407 &   407 \\
I05S08V10$^{\dagger}$ & 26.96 & $3.807\times10^6$ &  1 &  1 & 1.0 &  1 &   886 &   886 \\
I05S08V15             & 25.45 & $6.182\times10^6$ &  6 & 22 & 3.7 &  8 & 27220 & 12670 \\
I05S08V20             & 24.45 & $9.226\times10^6$ &  9 & 14 & 1.6 &  4 & 52540 & 42250 \\
I05S08V25             & 23.71 & $1.288\times10^7$ &  6 & 10 & 1.7 &  4 & 25110 & 10030 \\
I05S08V30             & 23.05 & $1.571\times10^7$ &  3 &  3 & 1.0 &  1 &  2945 &  1017 \\
I06S08V00             & 29.03 & $4.698\times10^5$ &  1 &  1 & 1.0 &  1 &   432 &   432 \\
I06S08V10$^{\dagger}$ & 25.01 & $1.079\times10^6$ &  1 &  1 & 1.0 &  1 &   451 &   451 \\
I06S08V15$^{\dagger}$ & 23.97 & $3.465\times10^6$ &  2 &  3 & 1.5 &  2 &  6434 &  5496 \\
I06S08V20             & 21.73 & $9.533\times10^6$ &  4 &  4 & 1.0 &  1 &  3187 &  1385 \\
I06S08V25             & 20.62 & $1.222\times10^7$ &  3 &  3 & 1.0 &  1 &  1713 &   703 \\
I06S08V30             & 19.63 & $6.869\times10^7$ &  8 & 21 & 2.6 & 10 & 74230 & 35270 \\
I07S08V00$^{\dagger}$ & 28.64 & $1.106\times10^6$ &  2 &  3 & 1.5 &  2 &   707 &   350 \\
I07S08V10             & 26.79 & $2.514\times10^6$ &  2 &  2 & 1.0 &  1 &  5918 &  5047 \\
I07S08V15             & 21.61 & $2.030\times10^7$ &  5 &  9 & 1.8 &  3 &  9004 &  3846 \\
I07S08V20             & 23.21 & $1.731\times10^7$ & 10 & 20 & 2.0 & 10 & 17330 &  3651 \\
I07S08V25             & 22.55 & $1.822\times10^7$ & 22 & 45 & 2.1 & 10 & 15260 &  1669 \\
I07S08V30$^{\dagger}$ & 22.57 & $1.705\times10^7$ &  1 &  2 & 2.0 &  2 &  2087 &  1106 \\
I08S08V00             & 28.42 & $2.637\times10^6$ &  1 &  1 & 1.0 &  1 & 14810 & 14810 \\
I08S08V10             & 27.91 & $2.786\times10^6$ &  1 &  1 & 1.0 &  1 &   939 &   939 \\
I08S08V15             & 25.85 & $8.700\times10^6$ &  4 &  8 & 2.0 &  5 & 45460 & 38290 \\
I08S08V20             & 25.01 & $1.543\times10^7$ & 21 & 44 & 2.1 &  6 & 41070 &  5501 \\
I08S08V25             & 24.31 & $1.980\times10^7$ &  1 &  1 & 1.0 &  1 & 15250 & 15250 \\
I08S08V30             & 23.55 & $2.587\times10^7$ &  2 &  3 & 1.5 &  2 &  1753 &   667 \\
\hline
\end{tabular}
\begin{flushleft}
{\it Notes.} Column 1: Model name.
Column 2: redshift ($z$).
Column 3: total halo mass ($\Mvir$) at the virial scale.
Columns 4 and 5: numbers of clouds ($\Ncore$) and stars ($\Nstar$).
Column 6: multiplicity of stars per cloud ($\Nstar/\Ncore$).
Column 7: maximum number of stars contained within a cloud ($\Nstarcloud$).
Columns 8 and 9: total and maximum mass of stars ($\Mstartot$ and $\Mstarmax$).
For models in which HD cooling is effective, the model names in the first column are marked with $\dagger$.
\end{flushleft}
\end{table*}

\begin{table*}
\centering
\contcaption{
}
\begin{tabular}{lccccccccc}
\hline
Model &
$z$ &
$\Mvir$ &
$\Ncore$ &
$\Nstar$ &
$\Nstar/\Ncore$ &
$\Nstarcloud$ &
$\Mstartot$ &
$\Mstarmax$ \\
&
&
($\msun$) &
&
&
&
&
($\msun$) &
($\msun$) \\
\hline
I09S08V00             & 28.03 & $2.503\times10^6$ &  1 &  1 & 1.0 &  1 & 13840 & 13840 \\
I09S08V10             & 29.63 & $1.061\times10^6$ &  1 &  1 & 1.0 &  1 &  1070 &  1070 \\
I09S08V15             & 26.45 & $4.638\times10^6$ &  8 & 14 & 1.8 &  5 & 13270 &  5332 \\
I09S08V20             & 25.16 & $7.538\times10^6$ &  4 &  8 & 2.0 &  3 & 10170 &  3496 \\
I09S08V25             & 24.21 & $1.002\times10^7$ &  2 &  2 & 1.0 &  1 &  1074 &   605 \\
I09S08V30             & 23.48 & $1.304\times10^7$ &  1 &  2 & 2.0 &  2 &  5297 &  4448 \\
I10S08V00$^{\dagger}$ & 27.61 & $9.544\times10^5$ &  1 &  2 & 2.0 &  2 &   840 &   449 \\
I10S08V10             & 25.38 & $2.608\times10^6$ &  4 &  5 & 1.3 &  2 & 10140 &  9535 \\
I10S08V15             & 24.20 & $3.801\times10^6$ &  1 &  1 & 1.0 &  1 &   762 &   762 \\
I10S08V20$^{\dagger}$ & 23.04 & $6.784\times10^6$ &  4 &  4 & 1.0 &  1 &  1224 &   756 \\
I10S08V25             & 21.37 & $2.009\times10^7$ &  7 & 12 & 1.7 &  6 & 14960 &  2522 \\
I10S08V30             & 20.79 & $2.721\times10^7$ & 16 & 34 & 2.1 & 14 & 86240 & 13340 \\
I11S08V00             & 27.60 & $2.675\times10^6$ &  1 &  1 & 1.0 &  1 &   947 &   947 \\
I11S08V10             & 25.42 & $5.918\times10^6$ &  2 &  3 & 1.5 &  2 &  1907 &  1022 \\
I11S08V15             & 24.51 & $1.113\times10^7$ &  3 &  5 & 1.7 &  3 & 38720 & 23130 \\
I11S08V20             & 23.75 & $1.343\times10^7$ &  2 &  2 & 1.0 &  1 &  1397 &   823 \\
I11S08V25             & 21.09 & $4.029\times10^7$ & 22 & 32 & 1.5 &  4 & 35190 & 20810 \\
I11S08V30             & 20.85 & $4.055\times10^7$ & 26 & 36 & 1.4 &  3 & 25830 &  6838 \\
I12S08V00$^{\dagger}$ & 26.61 & $6.222\times10^5$ &  1 &  3 & 3.0 &  3 &  1182 &   420 \\
I12S08V10$^{\dagger}$ & 22.58 & $2.168\times10^6$ &  1 &  1 & 1.0 &  1 &   246 &   246 \\
I12S08V15             & 20.31 & $6.683\times10^6$ &  1 &  1 & 1.0 &  1 &  1043 &  1043 \\
I12S08V20             & 19.09 & $1.027\times10^7$ &  6 & 10 & 1.7 &  4 & 98050 & 49040 \\
I12S08V25             & 18.67 & $1.052\times10^7$ &  3 &  4 & 1.3 &  2 & 21500 & 18550 \\
I12S08V30             & 18.13 & $1.242\times10^7$ &  3 &  4 & 1.3 &  2 &  8475 &  3293 \\
I13S08V00             & 26.43 & $7.845\times10^5$ &  1 &  1 & 1.0 &  1 &  1006 &  1006 \\
I13S08V10             & 23.46 & $2.268\times10^6$ &  3 &  3 & 1.0 &  1 &  5391 &  3426 \\
I13S08V15             & 21.70 & $4.545\times10^6$ &  2 &  2 & 1.0 &  1 & 14810 &  8388 \\
I13S08V20             & 19.55 & $7.821\times10^6$ &  1 &  1 & 1.0 &  1 &  1256 &  1256 \\
I13S08V25             & 17.93 & $1.259\times10^7$ &  2 &  2 & 1.0 &  1 &  6129 &  5672 \\
I13S08V30             & 17.23 & $1.652\times10^7$ &  1 &  1 & 1.0 &  1 &  1997 &  1997 \\
I14S08V00             & 25.42 & $1.471\times10^6$ &  2 &  2 & 1.0 &  1 &  2489 &  2002 \\
I14S08V10             & 22.70 & $4.561\times10^6$ &  5 &  9 & 1.8 &  3 & 52360 & 13330 \\
I14S08V15             & 22.21 & $5.590\times10^6$ &  5 & 10 & 2.0 &  5 & 25000 & 11230 \\
I14S08V20             & 21.27 & $7.451\times10^6$ &  1 &  2 & 2.0 &  2 & 22010 & 18720 \\
I14S08V25             & 20.05 & $1.134\times10^7$ &  1 &  1 & 1.0 &  1 &  1189 &  1189 \\
I14S08V30             & 19.79 & $1.468\times10^7$ &  2 &  2 & 1.0 &  1 &  2875 &  2369 \\
I15S08V00             & 24.71 & $1.016\times10^6$ &  2 &  2 & 1.0 &  1 &  1608 &   841 \\
I15S08V10             & 24.67 & $3.372\times10^6$ &  2 &  4 & 2.0 &  2 &  9809 &  8783 \\
I15S08V15             & 19.74 & $2.000\times10^7$ &  2 &  3 & 1.5 &  2 &  3159 &  1154 \\
I15S08V20             & 20.58 & $1.615\times10^7$ &  1 &  2 & 2.0 &  2 & 10800 & 10500 \\
I15S08V25             & 20.22 & $2.315\times10^7$ &  1 &  1 & 1.0 &  1 &  4747 &  4747 \\
I15S08V30             & 20.39 & $3.070\times10^7$ &  6 & 12 & 2.0 &  7 & 55160 & 14680 \\
I16S08V00             & 23.82 & $6.588\times10^5$ &  1 &  1 & 1.0 &  1 &   875 &   875 \\
I16S08V10             & 21.55 & $1.797\times10^6$ &  2 &  3 & 1.5 &  2 &  3898 &  3709 \\
I16S08V15             & 18.98 & $7.158\times10^6$ &  1 &  4 & 4.0 &  4 & 32000 & 29118 \\
I16S08V20             & 17.43 & $2.024\times10^7$ &  4 &  4 & 1.0 &  1 &  9045 &  7261 \\
I16S08V25             & 16.96 & $2.577\times10^7$ &  6 & 10 & 1.7 &  5 & 25690 & 21720 \\
I16S08V30             & 16.73 & $2.520\times10^7$ &  1 &  3 & 3.0 &  3 &  9265 &  8217 \\
I17S08V00             & 22.79 & $3.841\times10^5$ &  1 &  1 & 1.0 &  1 &   799 &   799 \\
I17S08V10             & 18.06 & $5.876\times10^6$ &  1 &  1 & 1.0 &  1 &  1206 &  1206 \\
I17S08V15             & 17.50 & $6.021\times10^6$ &  2 &  2 & 1.0 &  1 &  2445 &  1359 \\
I17S08V20             & 17.72 & $4.470\times10^6$ &  1 &  1 & 1.0 &  1 &   843 &   843 \\
I17S08V25             & 17.25 & $5.552\times10^6$ &  4 &  5 & 1.3 &  2 & 11510 &  5245 \\
I17S08V30             & 16.96 & $5.597\times10^6$ &  3 &  5 & 1.7 &  3 &  2473 &  1046 \\
 
\hline
\end{tabular}
\end{table*}

\begin{table*}
\centering
\contcaption{
}
\begin{tabular}{lccccccccc}
\hline
Model &
$z$ &
$\Mvir$ &
$\Ncore$ &
$\Nstar$ &
$\Nstar/\Ncore$ &
$\Nstarcloud$ &
$\Mstartot$ &
$\Mstarmax$ \\
&
&
($\msun$) &
&
&
&
&
($\msun$) &
($\msun$) \\
\hline
I18S08V00$^{\dagger}$ & 22.02 & $3.695\times10^6$ &  1 &  1 & 1.0 &  1 &   399 &   399 \\
I18S08V10$^{\dagger}$ & 22.30 & $2.768\times10^6$ &  1 &  2 & 2.0 &  2 &  4361 &  3122 \\
I18S08V15$^{\dagger}$ & 21.27 & $3.585\times10^6$ &  2 &  5 & 2.5 &  4 &  4773 &  3186 \\
I18S08V20             & 19.33 & $1.058\times10^7$ &  6 & 10 & 1.7 &  4 &  5559 &  2729 \\
I18S08V25             & 18.23 & $1.702\times10^7$ &  3 &  4 & 1.3 &  2 & 31630 & 21350 \\
I18S08V30             & 18.02 & $1.768\times10^7$ &  3 &  3 & 1.0 &  1 &  1667 &   667 \\
I19S08V00$^{\dagger}$ & 21.08 & $1.602\times10^6$ &  1 &  1 & 1.0 &  1 &  1117 &  1117 \\
I19S08V10$^{\dagger}$ & 20.98 & $1.563\times10^6$ &  1 &  1 & 1.0 &  1 &   469 &   469 \\
I19S08V15             & 19.19 & $2.700\times10^6$ &  3 &  3 & 1.0 &  1 &  1676 &  1099 \\
I19S08V20             & 17.33 & $9.673\times10^6$ &  3 &  3 & 1.0 &  1 &  1361 &   855 \\
I19S08V25             & 17.01 & $1.038\times10^7$ &  1 &  1 & 1.0 &  1 &   936 &   936 \\
I19S08V30             & 16.00 & $1.360\times10^7$ &  2 &  2 & 1.0 &  1 &   729 &   503 \\
I20S08V00             & 16.52 & $8.400\times10^6$ &  1 &  3 & 3.0 &  3 & 12480 &  5862 \\
I20S08V10$^{\dagger}$ & 17.74 & $3.178\times10^6$ &  1 &  1 & 1.0 &  1 &   756 &   756 \\
I20S08V15             & 16.86 & $4.375\times10^6$ &  1 &  4 & 4.0 &  4 & 19200 & 12130 \\
I20S08V20             & 15.61 & $5.969\times10^6$ &  2 &  3 & 1.5 &  2 & 25260 & 23210 \\
I20S08V25             & 15.79 & $6.027\times10^6$ &  1 &  1 & 1.0 &  1 &   527 &   527 \\
I20S08V30             & 16.49 & $4.599\times10^6$ &  1 &  2 & 2.0 &  2 &   709 &   427 \\
\hline
\end{tabular}
\end{table*}

\begin{table*}
\centering
\caption{Results of new 18 models.}
\label{table:new18models}
\begin{tabular}{lccccccccc}
\hline
Model &
$z$ &
$\Mvir$ &
$\Ncore$ &
$\Nstar$ &
$\Nstar/\Ncore$ &
$\Nstarcloud$ &
$\Mstartot$ &
$\Mstarmax$ \\
&
&
($\msun$) &
&
&
&
&
($\msun$) &
($\msun$) \\
\hline
I02S10V00             & 42.09 & $1.523\times10^6$ &  1 &  1 & 1.0 &  1 &   796 &   796 \\
I02S10V30             & 33.45 & $6.576\times10^6$ &  2 &  3 & 1.5 &  2 & 14150 & 13800 \\
I02S11V00             & 40.65 & $2.013\times10^6$ &  1 &  1 & 1.0 &  1 &  4035 &  4035 \\
I02S11V30             & 36.96 & $6.789\times10^6$ &  2 &  2 & 1.0 &  1 & 17200 & 16960 \\
I02S12V00             & 49.24 & $2.229\times10^6$ &  3 &  3 & 1.0 &  1 & 19910 & 18970 \\
I02S12V30             & 40.36 & $7.387\times10^6$ &  2 &  3 & 1.5 &  2 & 28620 & 28580 \\
I04S10V00             & 36.26 & $9.825\times10^5$ &  2 &  3 & 1.5 &  2 &  1559 &   804 \\
I04S10V30             & 28.80 & $8.782\times10^6$ &  1 &  3 & 3.0 &  3 & 22160 & 13130 \\
I04S11V00             & 39.60 & $1.022\times10^6$ &  4 &  7 & 1.8 &  4 &  8976 &  4640 \\
I04S11V30             & 32.05 & $8.519\times10^6$ &  2 &  2 & 1.0 &  1 &  1981 &  1233 \\
I04S12V00             & 42.79 & $1.159\times10^6$ &  4 &  7 & 1.8 &  4 &  7792 &  2840 \\
I04S12V30             & 35.16 & $9.120\times10^6$ &  1 &  2 & 2.0 &  2 &  7356 &  6351 \\
I16S10V00             & 29.08 & $3.440\times10^5$ &  1 &  1 & 1.0 &  1 &   699 &   699 \\
I16S10V30             & 20.28 & $2.668\times10^7$ &  2 &  4 & 2.0 &  2 & 17890 &  8192 \\
I16S11V00             & 32.32 & $3.276\times10^5$ &  1 &  1 & 1.0 &  1 &   700 &   700 \\
I16S11V30             & 22.42 & $2.919\times10^7$ &  4 &  5 & 1.3 &  2 &  6164 &  1649 \\
I16S12V00             & 35.84 & $2.683\times10^5$ &  1 &  1 & 1.0 &  1 &   543 &   543 \\
I16S12V30             & 24.64 & $2.826\times10^7$ &  5 &  7 & 1.4 &  2 &  6542 &  2595 \\
\hline
\end{tabular}
\begin{flushleft}
{\it Notes.} Columns are the same as in Table~\ref{table:120models} for the additional 18 models (see Section~\ref{sec:methods_additional-models}).
\end{flushleft}
\end{table*}

\section{Coarse-bin summary of multiplicity and stellar-mass statistics}
\label{sec:app:coarse}

To complement the trends discussed in the main text, we also summarize the model statistics after combining the haloes into three broader streaming-velocity regimes: low ($\vsvsigma=0$--$1$), intermediate ($1.5$--$2$), and high ($2.5$--$3$).
This coarse binning suppresses halo-to-halo scatter and provides a more statistically robust summary of the overall environmental trends than the finer sampling shown in the main text.
In the table labels below, GH, GM, and GL denote the three halo classes ({\it High}, {\it Middle}, and {\it Low}), while the suffixes 1, 2, and 3 denote the low-, intermediate-, and high-streaming bins, respectively.
Thus, for example, GH1 corresponds to the {\it High} class with $\vsvsigma=0$--$1$, whereas GL3 corresponds to the {\it Low} class with $\vsvsigma=2.5$--$3$.

Table~\ref{table:group_class-vs_mean} provides a compact reference summary of the multiplicity trends discussed in Section~\ref{sec:res:Vsv-multiplicity}.
In particular, it confirms that low-streaming groups are more often dominated by single-star haloes, whereas intermediate- and high-streaming groups tend to host larger numbers of Jeans-unstable clouds and first stars, leading to higher fractions of FSCs.

Table~\ref{table:group_class-vs_imf} provides the corresponding coarse-bin summary of the stellar-mass statistics discussed in Section~\ref{sec:res:imf} and shown in Figures~\ref{fig:Vsv-Mstarmax} and \ref{fig:Vsv-fmassbin}.
It confirms that the $10$--$40\,\msun$ fraction remains small across all environments, while the PISN range and the very-high-mass tail retain non-negligible contributions in many intermediate- and high-streaming groups.

\begin{table*}
\centering
\caption{Mean properties and fraction of haloes grouped by $z-\Mvir$ class and streaming velocity.}
\label{table:group_class-vs_mean}
\begin{tabular}{lcccccccccccccc}
\hline
Group &
$\nmodel$ &
$\langle z \rangle$ &
$\langle \Mvir \rangle$ &
$\langle \Ncore \rangle$ &
$\langle \Nstar \rangle$ &
$\langle \Nstar/\Ncore \rangle$ &
$\langle \Nstarcloud \rangle$ &
$\langle \Mstartot \rangle$ &
$\langle \Mstarmax \rangle$ &
$f_{\rm single}$ &
$f_{\rm multi}$ &
$f_{\rm cl,5}$ &
$f_{\rm cl,10}$ \\
&
&
&
($\msun$) &
&
&
&
&
($\msun$) &
($\msun$) &
(\%)&
(\%)&
(\%)&
(\%)\\
\hline
GH1 & 10 & 30.03 & $2.26\times10^6$ & 1.30 &  1.70 & 1.20 & 1.40 &  2778 & 2499 & 70 &  30 &  0 &  0 \\
GH2 & 10 & 26.45 & $7.87\times10^6$ & 5.60 & 10.30 & 1.78 & 3.40 &  9935 & 5180 &  0 & 100 & 70 & 20 \\
GH3 & 10 & 22.67 & $3.15\times10^7$ & 7.00 & 10.60 & 1.73 & 3.00 & 10660 & 6172 & 10 &  90 & 40 & 30 \\
GM1 & 16 & 26.81 & $1.66\times10^6$ & 2.12 &  3.06 & 1.32 & 1.75 &  1952 & 1398 & 31 &  69 & 19 &  6 \\
GM2 & 16 & 22.87 & $8.82\times10^6$ & 4.25 &  7.88 & 1.66 & 3.44 &  6010 & 3560 & 19 &  81 & 44 & 31 \\
GM3 & 16 & 21.35 & $1.86\times10^7$ & 5.76 & 10.12 & 1.46 & 3.88 &  8238 & 3688 & 17 &  83 & 47 & 41 \\
GL1 & 14 & 21.85 & $1.97\times10^6$ & 1.21 &  1.64 & 1.39 & 1.43 &  1312 & 1088 & 64 &  36 &  0 &  0 \\
GL2 & 14 & 18.71 & $7.00\times10^6$ & 2.50 &  3.79 & 1.67 & 2.14 &  5644 & 4032 & 21 &  79 & 21 & 14 \\
GL3 & 14 & 17.24 & $1.23\times10^7$ & 2.43 &  3.36 & 1.40 & 1.93 &  3921 & 2688 & 21 &  79 & 21 &  7 \\
\hline
\end{tabular}
\begin{flushleft}
{\it Notes.}
Column 1: Group labels encode the three halo classes (H/M/L) and the streaming-velocity bin (1/2/3).
Classes {\it High}, {\it Middle}, and {\it Low} correspond to (I01, I02, I08, I09, I11), (I03--07, I10, I14, I15), and (I12, I13, I16--20), respectively \citep{Hirano2025FSC2}.
Bins 1--3 correspond to $\vsvsigma = 0$--1, 1.5--2, and 2.5--3, respectively.
We use models with $\sigma_8 = 0.83$ (S08; Table~\ref{table:120models}).
Column 2: number of haloes.
Column 3: redshift ($z$).
Column 4: total halo mass ($\Mvir$) at the virial scale.
Columns 5 and 6: numbers of clouds ($\Ncore$) and stars ($\Nstar$).
Column 7: multiplicity of stars per cloud ($\Nstar/\Ncore$).
Column 8: maximum number of stars contained within a cloud ($\Nstarcloud$).
Columns 9 and 10: total and maximum mass of stars ($\Mstartot$ and $\Mstarmax$).
Columns 11-14: $f_{\rm single}$, $f_{\rm multi}$, $f_{\rm cl,5}$, and $f_{\rm cl,10}$ denote the fractions of haloes hosting $\Nstar = 1$, $\Nstar \ge 2$, $\Nstar \ge 5$, and $\Nstar \ge 10$ first stars, respectively.
All halo-averaged quantities are computed as arithmetic means, except for the masses, which are geometric means.
\end{flushleft}
\end{table*}

\begin{table*}
\centering
\caption{Fraction of haloes grouped by $z-\Mvir$ class and streaming velocity.}
\label{table:group_class-vs_imf}
\begin{tabular}{lcccccccc}
\hline
Group &
$f_{<10}$ &
$f_{10-40}$ &
$f_{40-60}$ &
$f_{60-120}$ &
$f_{120-260}$ &
$f_{260-10^3}$ &
$f_{10^3-10^4}$ &
$f_{>10^4}$ \\
\hline
GH1 & 0     & 0     & 0     & 0.025 & 0.058 & 0.392 & 0.325 & 0.200 \\
GH2 & 0.036 & 0.110 & 0.009 & 0.040 & 0.153 & 0.445 & 0.112 & 0.094 \\
GH3 & 0.030 & 0.030 & 0.015 & 0.044 & 0.048 & 0.422 & 0.271 & 0.141 \\
GM1 & 0.034 & 0.019 & 0.007 & 0.053 & 0.114 & 0.624 & 0.110 & 0.040 \\
GM2 & 0.036 & 0.071 & 0.030 & 0.061 & 0.078 & 0.449 & 0.188 & 0.085 \\
GM3 & 0.007 & 0.035 & 0.007 & 0.039 & 0.076 & 0.303 & 0.498 & 0.036 \\
GL1 & 0     & 0     & 0     & 0.024 & 0.095 & 0.476 & 0.405 & 0     \\
GL2 & 0     & 0.018 & 0.029 & 0.071 & 0.083 & 0.240 & 0.466 & 0.074 \\
GL3 & 0.007 & 0.074 & 0.007 & 0.036 & 0.061 & 0.436 & 0.337 & 0.043 \\
\hline
\end{tabular}
\begin{flushleft}
{\it Notes.}
Group labels encode the combination of the halo class and the streaming-velocity range (see Table~\ref{table:group_class-vs_mean}).
$f_{M_1-M_2}$ denotes the fraction of first stars with masses in the range $M_1 \le \Mstar/\msun < M_2$.
$f_{<10}$ and $f_{>10^4}$ denote the fraction with $\Mstar < 10\,\msun$ and $\Mstar \ge 10^4\,\msun$.
\end{flushleft}
\end{table*}

\section{Bin-width dependence of Figure~11}
\label{app:binwidth_test}

\begin{figure}
\centering
\includegraphics[width=1.0\linewidth]{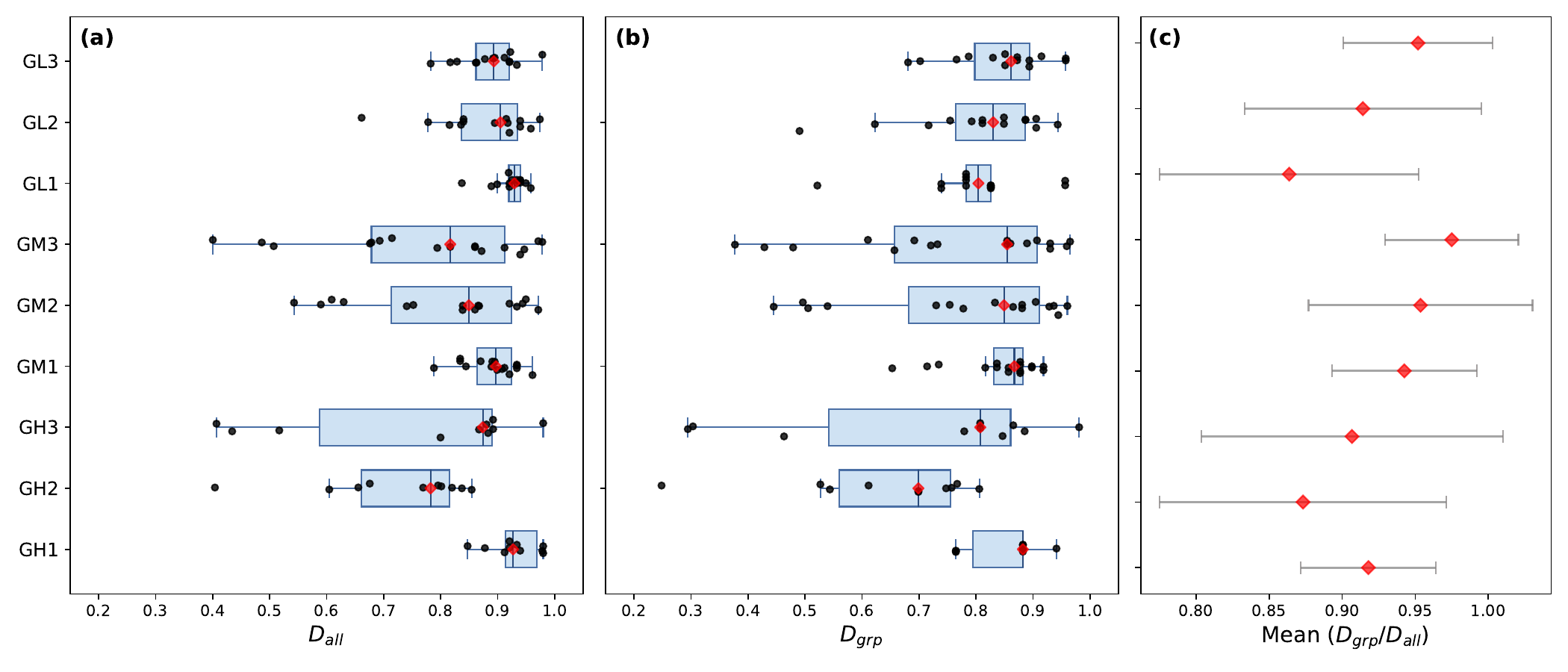}\\
\includegraphics[width=1.0\linewidth]{./figures/f_imf-dist_tv.pdf}\\
\includegraphics[width=1.0\linewidth]{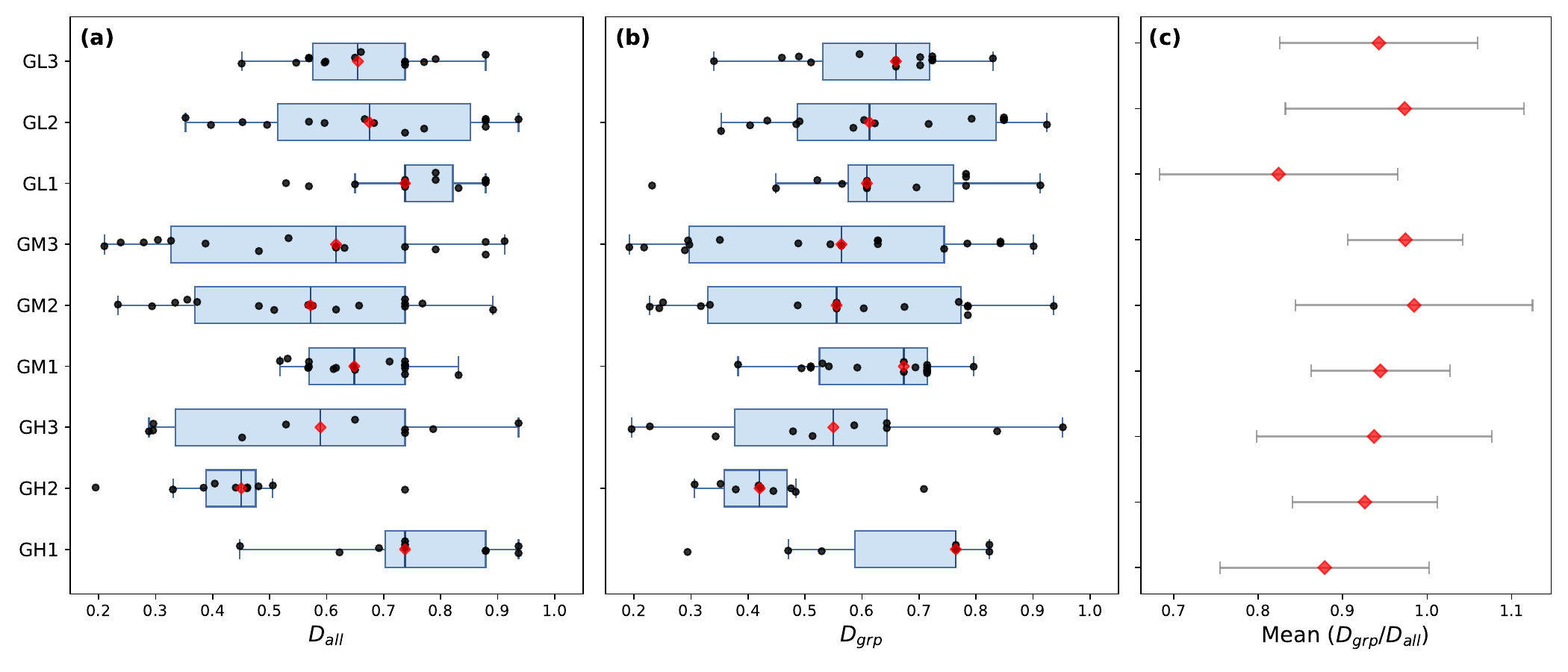}
\caption{
Comparison of IMF-distance results for three mass-bin widths of $0.5 \times$ (top), $1 \times$ (middle; Figure~\ref{fig:imf-dist}), and $2 \times$ (bottom).
The qualitative group-level trends are robust, while absolute distances are systematically smaller for coarser bins.
}
\label{fig:imf-dist-binwidth}
\end{figure}

To assess bin-width dependence, we recomputed the IMF-distance statistics using mass-bin widths of $0.5 \times$, $1 \times$, and $2 \times$.
The qualitative ordering and group-to-group trends in $D_{\rm all}$, $D_{\rm grp}$, and $D_{\rm grp}/D_{\rm all}$ are preserved across these choices.
As expected for coarse-grained histograms, the absolute values of $D_{\rm all}$ and $D_{\rm grp}$ decrease for larger bin widths.

\section{Datasets of Figure~12}
\label{app:Figure11}

Figure~\ref{fig:z-Mhalo-Mbh} includes, in addition to the 138 FSC haloes presented in this work, comparison points from previous simulations and observations.
In panel (a), observed galaxies are plotted at $M_{\rm halo} = 100\,M_{\ast}$ as a fiducial mapping assuming a stellar-to-halo mass ratio of $\sim 0.01$.
Observed redshifts are spectroscopic unless otherwise noted; open diamonds indicate photometric redshift estimates.
The orange hatched region shows MW-like progenitor merger trees from a semi-analytic model based on a large-volume $N$-body simulation \citep[Phi-4096;][]{IshiyamaHirano2025}.
It is constructed from 31 haloes with $\Mvir(z=0) \ge 5\times10^{11}\,\msun$.
In panel (b), observational points show central BH masses unless otherwise noted (LAP1-B is plotted at the inferred total Pop~III stellar mass).

\begin{table*}
\centering
\caption{Simulation datasets used in Figure~\ref{fig:z-Mhalo-Mbh} and their legend symbols.}
\label{tab:fig11_sim}
\begin{tabular}{lllll}
\hline
Label & Symbol & Reference & $\vsvsigma$ & Stellar-mass estimate \\
\hline
FSC (this work) & circle          & this work              & 0--3 & inferred from gas infall rate \\
H17             & square          & \citet{Hirano2017smbh} & 3    & accretion-phase simulation    \\
H15             & $\times$ marker & \citet{Hirano2015}     & 0    & inferred from gas infall rate \\
H14             & filled dot      & \citet{Hirano2014}     & 0    & accretion-phase simulation    \\
\hline
\end{tabular}
\end{table*}
\begin{table*}
\centering
\caption{Observational datasets plotted in panel (a) of Figure~\ref{fig:z-Mhalo-Mbh}.}
\label{tab:fig11_obs_a}
\begin{tabular}{ll}
\hline
Label & Objects and reference(s) \\
\hline
galaxy    & galaxy sample \citep{Perez-Gonzalez2025, NakazatoFerrara2025} \\
PG25      & PG25 sample \citep{Perez-Gonzalez2025} \\
SMBH host & UHZ1 \citep{Goulding2023_UHZ1}, GHZ9 \citep{Kovacs2024_GHZ9,Napolitano2025_GHZ9}, CEERS\_1019 \citep{Larson2203_CEERS_1019} \\
          & Abell2744-QSO1 \citep{Furtak2024_Abell2744-QSO1} \\
LRD host  & CAPERS-LRD-z9 \citep{Taylor2025_CAPERS-LRD-z9}, CANUCS-LRD-z8.6 \citep{Tripodi2025_CANUCS-LRD-z8.6}, UNCOVER-ID20466 \citep{Kokorev2023_UNCOVER-ID20466,Jones2026_UNCOVER-ID20466} \\
          & Mom-BH$\ast$-1 \citep{Naidu2025_Mom-BHstar-1}, Virgil \citep{Rinaldi2025_Virgil}, COS-66964 \citep{Akins2025_COS-66964} \\
R26       & LRD host sample \citep{Rusakov2026} \\
(N-rich)  & nitrogen-enhanced galaxy candidates \citep{Zhu2026} \\
LAP1-B    & LAP1-B \citep{Nakajima2025_LAP1-B} \\
MW        & Milky Way \\
\hline
\end{tabular}
\end{table*}
\begin{table*}
\centering
\caption{Observational datasets plotted in panel (b) of Figure~\ref{fig:z-Mhalo-Mbh}.}
\label{tab:fig11_obs_b}
\begin{tabular}{ll}
\hline
Label & Objects and reference(s) \\
\hline
SMBH         & (SMBH host in Table~\ref{tab:fig11_obs_a}), GNz-11 \citep{Maiolino2024_GNz-11}, C3D-z7AGN-z \citep{Lin2026_C3D-z7AGN-z}, ZS7 \citep{Ubler2024_ZS7} \\
             & J1342+0928 \citep{Banados2018_J1342+0928}, J1120+0641 \citep{Mortlock2011_J1120+0641}, Himiko-B \citep{Kiyota2025_Himiko-B}, J0100+2802 \citep{Wu2015_J0100+2802} \\
I20          & SMBH sample \citep{Inayoshi2020} \\
LRD          & (LRD host in Table~\ref{tab:fig11_obs_a}) \\
M24+R26      & LRD sample \citep{Matthee2024, Rusakov2026} \\
LAP1-B       & LAP1-B \citep{Nakajima2025_LAP1-B} \\
UMBH         & ultramassive BHs (mean) \citep{deNicola2025} \\
Sgr~A$^\ast$ & Sgr~A$^\ast$ \\
\hline
\end{tabular}
\end{table*}


\bsp
\label{lastpage}
\end{document}